\documentclass[%
 twocolumn,
superscriptaddress,
 amsmath,
 aps, pra,
]{revtex4-1}

\usepackage{color}
\usepackage{graphicx}
           
\renewcommand{\Re}{\mathop{\mathrm{Re}}\nolimits}           
   
\newcommand{\sign}{\mathop{\rm sign}} 

\newcommand{\grad}{\boldsymbol{\nabla}}
\newcommand{\diag}{\mathop{\rm diag}}

\begin{document}

\title{Signature of resonant modes in radiative heat current noise spectrum}

\author{Jonathan L. Wise}
\altaffiliation[Present address: ]{Univ. Bordeaux, CNRS, LOMA, UMR 5798, F-33400 Talence, France}
\affiliation{Univ. Grenoble Alpes, CNRS, LPMMC, 38000 Grenoble, France}

\author{Nathan Roubinowitz}
\affiliation{ENS Paris-Saclay, 91190 Gif-sur-Yvette, France}

\author{Wolfgang Belzig}
\affiliation{Fachbereich Physik, Universit\"at Konstanz, 78457 Konstanz, Germany}

\author{Denis M. Basko}
\affiliation{Univ. Grenoble Alpes, CNRS, LPMMC, 38000 Grenoble, France}

\begin{abstract}
Radiative heat transfer between bodies is often dominated by a narrow resonance in the transmission, e.~g. due to a cavity mode or a surface excitation. 
However, this resonant character is not visible in the average heat current. Here, we show that the noise spectrum of heat current can serve as a direct probe of the heat-carrying excitations. Namely, the resonant mode produces a sharp peak in the noise spectrum with a width related to the mode lifetime. We demonstrate that heat transfer in realistic superconducting circuits or between two-dimensional metals can realize our predictions.

\end{abstract}

\maketitle

\section{Introduction}

Noise is well known to contain additional useful information as compared to the average signal~\cite{Landauer1998}. This is well exploited in the field of electronic quantum transport, where the electric current noise is used to characterise conduction mechanisms and nature of charge carriers~\cite{Blanter2000}.
Here, we theoretically study the noise spectrum of energy current in near-field radiative heat transfer, and show that it provides valuable information on the nature of heat-carrying excitations, absent in the average energy current.
Energy current noise is much harder to probe experimentally than that of the electric current; however, the recent measurement of electronic temperature fluctuations, directly related to heat current noise~\cite{Karimi2020}, as well as the proposal for measuring high-frequency fluctuations~\cite{Karimi2021} suggest that such experiments are a matter of a near future.

Near-field thermal radiation can be dramatically enhanced with respect to the Planckian limit and may provide a dominant mechanism for heat transfer between two nearby bodies, isolated galvanically~\cite{Polder1971}. Over several decades, it has been studied in a wide variety of settings, both theoretically and experimentally, as described in several reviews~\cite{Joulain2005, Volokitin2007, Song2015, Bimonte2017, Biehs2021}. Using Rytov's framework of fluctuating electrodynamics \cite{Rytov1953, Rytov1989}, for the average power $\langle{P}\rangle$ transferred between two bodies held at temperatures $T_2>T_1$, one routinely finds a Landauer-type expression,
\begin{equation}
\langle P \rangle = \int_{0}^\infty \frac{d\omega}{\pi} \,\hbar \omega \, \mathcal{T}(\omega) \, \left[ \mathcal{N}(\omega,T_2) - \mathcal{N}(\omega,T_1)\right], 
\label{eq:Landauer}
\end{equation}
where $\mathcal{N}(\omega,T)\equiv1/(e^{\hbar\omega/T}-1)$ is the Bose-Einstein distribution at temperature~$T$ and we set the Boltzmann constant $k_\mathrm{B}=1$. The transmission function $\mathcal{T}(\omega)$ is determined by the system's geometry and the constituting materials. It encodes all information about the system's excitations which are responsible for the heat transfer.
If the heat is transferred by a broad continuum of excitations, then $\mathcal{T}(\omega)$ is a smooth function. Alternatively, $\mathcal{T}(\omega)$ may contain sharp resonances, e.~g., when the heat transfer is dominated by surface modes~\cite{Pendry1999, Yang2018}, or photons in a superconducting resonator~\cite{Meschke2006}. However, due to the frequency integration in Eq.~(\ref{eq:Landauer}) such information about the nature of heat-carrying excitations is not easy to extract from the average signal $\langle P\rangle$ which in all cases remains a smooth function of the temperatures~$T_1,T_2$ (typically, a power law). 

\begin{figure}
\centering
\includegraphics[width = 0.44\textwidth]{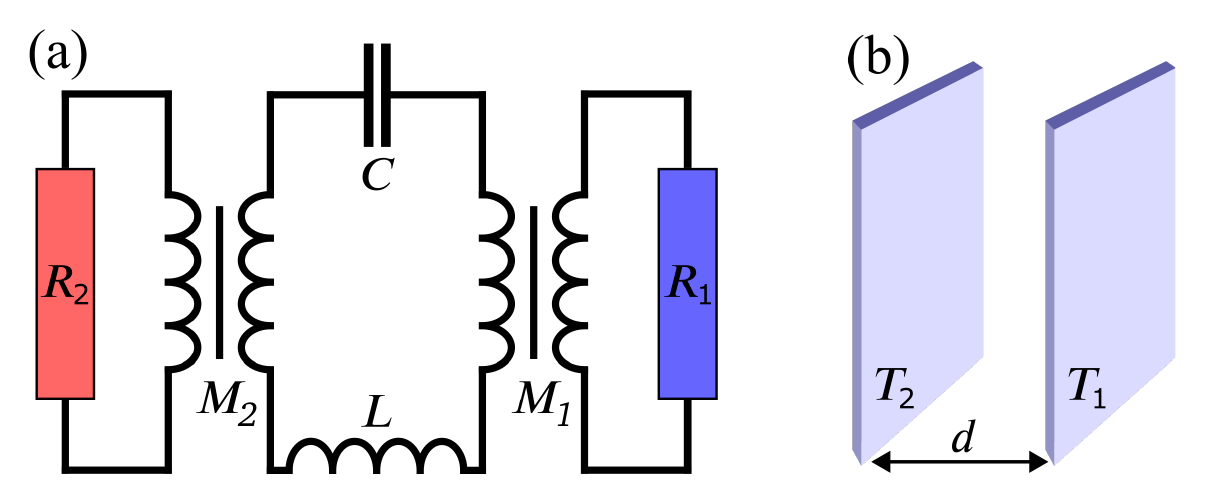}
\caption{Two examples of resonant heat transfer. (a) Electric circuit used to model heat transfer via a superconducting resonator mode: hot and cold resistors coupled to an $LC$ resonators via mutual inductances~$M$. (b)~Two-dimensional metallic sheets separated by a vacuum gap, $d$, exchanging heat due to strongly coupled surface plasmons.}
\label{fig:2systems}
\end{figure}

Here, we show that the information about resonances in $\mathcal{T}(\omega)$ can be recovered by studying the finite frequency heat current noise spectrum,
\begin{equation}\label{eq:0d_spec_def}
\mathcal{S}(\Omega) = \int_{-\infty}^\infty{d}t\,e^{i\Omega{t}}
\left[\frac12{\langle \{\hat{P}(t),\hat{P}(0)\}\rangle}-\langle P \rangle^2\right],
\end{equation}
where $\hat{P}(t)$ is the quantum-mechanical Heisenberg operator of the instantaneous power and $\{\ldots\}$ denotes the anticommutator.
Namely, $\mathcal{S}(\Omega)$ exhibits a sharp feature at $\Omega\to0$, whose width is related to the resonant mode lifetime. 
We illustrate this general conclusion by calculating the noise in the two example systems shown in Fig.~\ref{fig:2systems}. The first is a superconducting resonator that may be modelled by an effective quantum circuit, where the heat transfer takes place via the resonator mode~\cite{Pascal2011}. The second is a macroscopic system of two-dimensional (2D) metallic layers modeled by Drude conductivity, where the heat transfer is due to strongly coupled surface plasmon modes~\cite{Wise2020}. We find that the observability of the interesting feature in the noise spectrum depends on the timescales of the intrinsic temperature-relaxation dynamics of the material. We remark that radiative heat current fluctuation have been studied before with a focus either on the equal-time variance \cite{Biehs2018, Herz2020} or on low frequency noise via the full counting statistics \cite{Golubev2015, Tang2018}; here we show how a useful information about heat-carrying resonant modes can be extracted from the finite-frequency noise spectrum.



\section{Example 1: superconducting resonator}

In experiments of the type as in Ref.~\cite{Meschke2006}, metals in a micrometric setup are held at different temperatures and exchange heat via superconducting leads that form a resonator. The experiment is performed at sub-Kelvin temperatures such that the dominant heat transfer mechanism is photons of a single resonator mode. Since the system size is much smaller than the thermal wavelength of photons, such a setup may be modelled as an effective zero-dimensional electric circuit like that in Fig.~\ref{fig:2systems}~(a).
The normal metals are represented by resistors $R_1,R_2$ at temperatures $T_1<T_2$, and the superconducting resonator by an $LC$ circuit. The three elements are coupled via mutual inductances whose magnitudes~$M_1,M_2$ play the role of coupling constants. 

The average power $\langle{P}\rangle$ flowing between the two resistors in the circuit of Fig.~\ref{fig:2systems}~(a) has been calculated using the circuit version of the fluctuational electrodynamics~\cite{Pascal2011}. It is given by Eq.~\eqref{eq:Landauer} with~$\mathcal{T}(\omega) = 2\left |U_{12}(\omega) \right|^2/(R_1 R_2)$, where the coupling term is given by
\begin{equation}
U_{12}(\omega) = \frac{i \omega^3 C M_1M_2}{1-\omega^2 LC -i\omega^3 C (M_1^2/R_1 + M_2^2/R_2)}.
\label{eq:U_12}
\end{equation}
$U_{12}(\omega)$ has a complex pole which results in a Lorentzian resonance in $\mathcal{T}(\omega)$ at the resonator frequency $\omega_0\equiv  1/\sqrt{LC}$ with the full width of half maximum $\gamma = \gamma_1 + \gamma_2$, $\gamma_{1,2} = \omega_0^2M_{1,2}^2/(R_{1,2}L)$. Under the weak coupling assumption, $\gamma_{1,2}\ll \omega_0$, the contribution of this Lorentzian peak to the average heat current is $\langle P \rangle = (T_2-T_1)\gamma_1\gamma_2/\gamma$ (see Appendix~\ref{app:circuit}).
Moreover, for $\hbar\omega_0 \ll T_2 \ll \hbar\omega_0 (\omega_0/\gamma)^{1/3}$ the resonant contribution dominates the heat transfer.
Notice that although the simple expression for $\langle P \rangle$ does contain the properties of the resonance, the resonant character of $\mathcal{T}(\omega)$ cannot be deduced from the temperature dependence of $\langle P \rangle$ without an \textit{a~priori} knowledge.


To calculate the power noise spectrum $\mathcal{S}(\Omega)$, Eq.~(\ref{eq:0d_spec_def}), we adopt the standard approach and model each resistor as a bath of quantum harmonic oscillators, and the $LC$ contour as another harmonic oscillator (see Appendix~\ref{app:circuit} for details). The Hamiltonian of the inductive coupling between the oscillators is bilinear in the bosonic creation and annihilation operators, so the model is exactly solvable. 
We define the operator $\hat{P}(t)$ as the power dissipated in the resistor $R_1$ which is equivalent to the time derivative of the energy stored in the corresponding bath. This gives
\begin{equation}\label{eq:Pcircuit}
\hat{P}(t) = -\frac{1}{2}\left\{ \hat{I}_1(t), M_1\, \frac{d\hat{I}_{LC}(t)}{dt} \right\},
\end{equation}
where $\hat{I}_1(t)$ and $\hat{I}_{LC}(t)$ are the Heisenberg picture operators of the electric current in the resistor $R_1$ and in the $LC$ loop, respectively, so that $-M_1d\hat{I}_{LC}/dt$ is the voltage produced by the mutual inductance. $\hat{I}_1(t)$ and $\hat{I}_{LC}(t)$ may be found by solving the appropriate Heisenberg equations of motion, which take the form of quantum Langevin equations whose random forces are nothing but Johnson-Nyquist noise associated with each resistor.

Let us focus on the case of a large temperature bias $T_2\gg T_1$, when the noise spectrum feature is the most easy to measure, as discussed below. 
In this limit the dominant contribution to the heat current noise spectrum is given by the integral
\begin{align}
\mathcal{S}(\Omega) ={}&{}\int_{-\infty}^\infty \frac{d\omega}{2\pi} \left[1+\coth\frac{\hbar\omega}{2T_2}\coth\frac{\hbar(\Omega-\omega)}{2T_2}\right]  \nonumber \\
 {}&{}\times2\hbar^2\,\frac{\omega(\Omega-\omega)}{R_1^2 R_2^2}\left| U_{12}(\omega)\right|^2 \left|U_{12}(\Omega-\omega) \right|^2.
 \label{eq:spectrum_integral}
\end{align}
We note that the 1 in the square brackets, which ensures the integral convergence, can be recovered only from the quantum mechanical calculation, while a semi-classical treatment neglecting operator commutators misses this term. Regarding Eq.~\eqref{eq:spectrum_integral}, we notice that near the resonance, the product $\left|  U_{12}(\omega)\right|^2 \left|U_{12}(\Omega-\omega) \right|^2$ may be approximated by two Lorentzians which overlap if $|\Omega|\lesssim\gamma$.
Focusing on $\Omega \ll \omega_0 \ll T_2/\hbar$, we evaluate the frequency integral in this Lorentzian approximation (see Appendix~\ref{app:circuit}), and obtain 
\begin{equation}
\mathcal{S}(\Omega) = \frac{2\gamma}{\Omega^2 + \gamma^2}\,\left(\frac{\gamma_1 \gamma_2}{\gamma}\, T_2\right)^2, 
\label{eq:0d_res}
\end{equation}
the first main result of our work.
It shows that $\mathcal{S}(\Omega)$ exhibits a Lorentzian feature centred at $\Omega = 0$ whose width is determined by the damping rate of resonant photons in the circuit. This feature is inherited from the overlap of two Lorentzian factors and is a signature of the resonant nature of the heat transfer. Finally, we remark that $\mathcal{S}(\Omega\to0)$ has a form similar to that of electric current shot noise \cite{Blanter2000}: $\mathcal{S}(0) = 2\hbar\omega_0 \langle P \rangle \mathcal{F}$, where $\mathcal{F}=(T_2/\hbar\omega_0)(\gamma_1\gamma_2/\gamma^2)\gg 1$ is a large Fano factor indicating photon bunching, as expected for Bose statistics. 

\section{Example 2: two metallic layers}

The layers are assumed to be thin in comparison to the vacuum gap~$d$, separating them [Fig.~\ref{fig:2systems}~(b)], and to the skin depth. Modelled as degenerate 2D electron gases with impurities, the layers are assumed to be characterised by their 2D Drude conductivity, $\sigma(\omega)=\sigma_\mathrm{dc}/(1-i\omega\tau)$, and temperature-independent electron momentum relaxation time due to impurity scattering,~$\tau$, the same for both layers. We will focus on structures with $2\pi\sigma_\mathrm{dc}/c\lesssim1$ (which is typically the case for atomically thin 2D materials such as doped graphene or transition metal dichalcogenides); then the near-field heat transfer is dominated by the electrostatic Coulomb interaction between electrons~\cite{Wise2021}.

When the layers are held at different temperatures $T_2> T_1$, the average heat current per unit area $\langle J \rangle$ is given by an expression similar to Eq.~\eqref{eq:Landauer}, involving the transmission function $\mathcal{T}(\mathbf{k},\omega)$ which depends on the in-plane wavevector~$\mathbf{k}$, and also including integration over $d^2\mathbf{k}/(2\pi)^2$. For not too small $d\gg{v}_F\tau(2\pi\sigma_\mathrm{dc}/v_F)^{-1/2}$, with $v_F$ being the Fermi velocity, one can neglect the spatial dispersion of the conductivity, and then~\cite{Wise2020}
\begin{align}
&\mathcal{T}(\mathbf{k},\omega)=2\left[\Re\sigma(\omega)\right]^2|u_{12}(\mathbf{k},\omega)|^2,\\
&u_{12}(\mathbf{k},\omega)=\frac{2\pi{i}k\omega{e}^{-kd}}{[\omega+2\pi{i}k\sigma(\omega)]^2+[2\pi{k}\sigma(\omega)e^{-kd}]^2}.\label{eq:u12=}
\end{align}
This expression has poles in the complex plane of~$\omega$, which correspond to the underdamped (when $\omega\gg1/\tau$) strongly coupled (when $kd\lesssim1$) plasma oscillations of the two layers. In particular, the antisymmetric plasmon with the charge densities of the two layers oscillating out of phase, has a linear dispersion with the velocity $v_-=\sqrt{2\pi\sigma_\mathrm{dc}d/\tau}$. It was shown to dominate the average heat current $\langle{J}\rangle=\zeta(3)\,(T_2^3-T_1^3)/(8\pi^2\hbar^2\sigma_\mathrm{dc}d)$ for temperatures $\hbar/\tau\ll{T}_2\ll\hbar{v}_-/d$ (with $\zeta(x)$ denoting the Riemann $\zeta$~function)~\cite{Wise2020}. Again, one cannot deduce the resonant character of $\mathcal{T}(\mathbf{k},\omega)$ form the dependence of the average $\langle{J}\rangle$ on $T_1,T_2$ or~$d$.

\begin{figure}
\centering
\includegraphics[width = 0.48\textwidth]{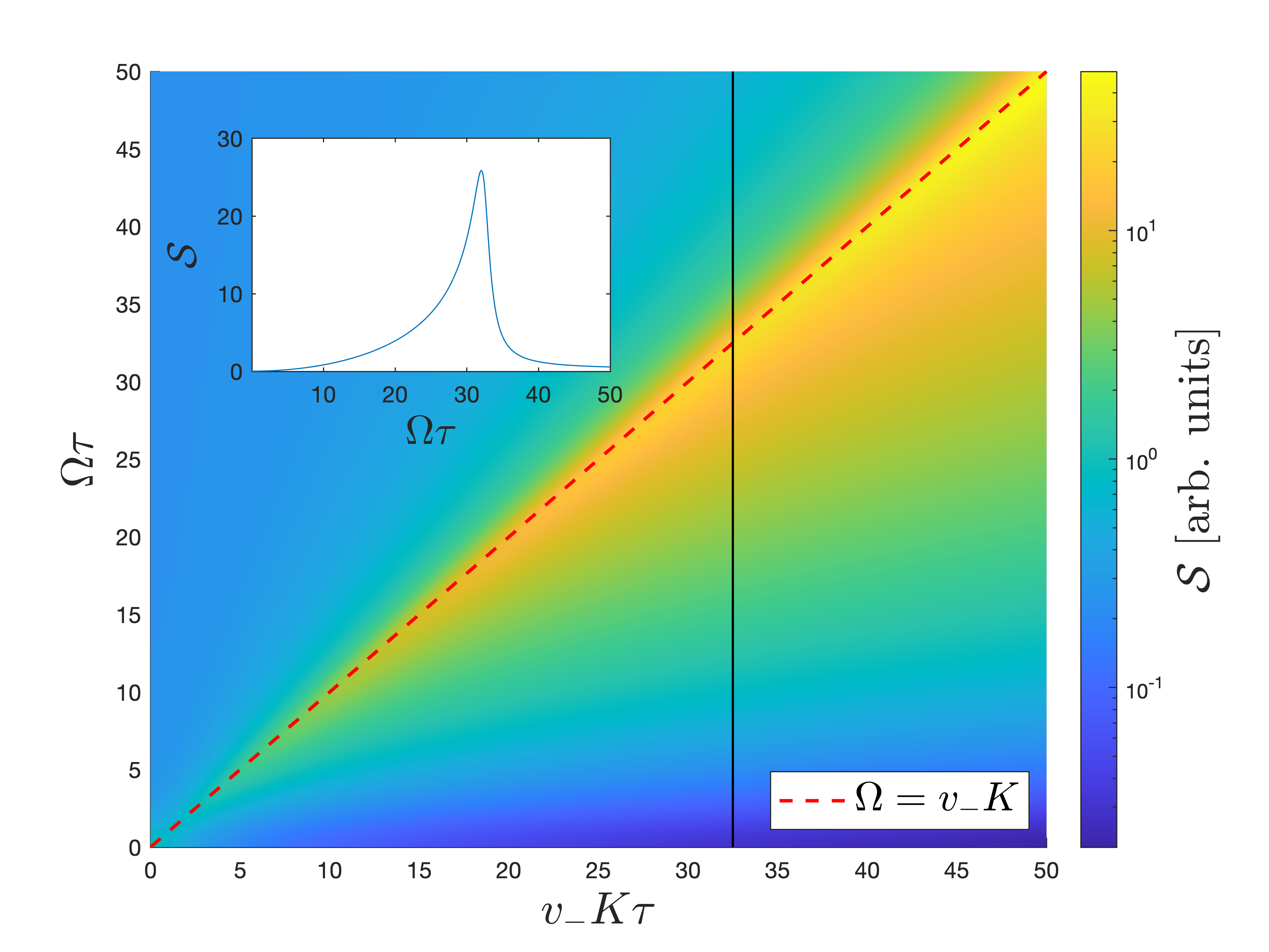}
\caption{Noise spectrum $\mathcal{S}(\mathbf{K},\Omega)$ of near-field radiative heat current between two-dimensional metallic layers, Eq.~\eqref{eq:2d_res}, as a function of the dimensionless variables $v_-K\tau$, $\Omega \tau$, in the regime when the heat transfer is dominated by antisymmetric surface plasmons. The dashed line shows the antisymmetric plasmon dispersion $\Omega = v_- K$. Inset: Cut at a fixed wavevector (solid black line) illustrating the shape of the feature, whose width is determined by the plasmon lifetime, $\tau$.}
\label{fig:2d_res}
\end{figure}

We now turn to the noise spectrum of the local heat current, defined as the power per unit area, dissipated at a given point~$\mathbf{r}$ of layer~1,
\begin{subequations}\begin{align}
&\hat{J}(\mathbf{r}) =\, :\hat{\mathbf{j}}_1(\mathbf{r})\cdot\hat{\mathbf{E}}_1(\mathbf{r}):\,,
\label{eqn:heatop}\\
&\hat{\mathbf{E}}_1(\mathbf{r})=
-\grad\int \left[\frac{\hat\rho_1(\mathbf{r}')}{|\mathbf{r}-\mathbf{r}'|}
+\frac{\hat\rho_2(\mathbf{r}')}{\sqrt{|\mathbf{r}-\mathbf{r}'|^2+d^2}}\right]d^2\mathbf{r'}, 
\end{align}\end{subequations}
where the operators $\hat{\mathbf{j}}_{1,2}(\mathbf{r})$ and $\hat{\rho}_{1,2}(\mathbf{r})$ are the 2D (surface) charge current and density, respectively, in layers 1 and~2, and $:\ldots:$ denotes the normal ordering of the electronic creation and annihilation operators which enter $\hat{\mathbf{j}}_{1,2}$ and $\hat{\rho}_{1,2}$. This is the standard expression for Joule losses due to electric field $\hat{\mathbf{E}}_1(\mathbf{r})$ produced in layer~1 by charges in the two layers (see Appendix~\ref{app:layers} for the formal derivation).
Since we are dealing with a spatially extended system, the natural object to characterise the noise spectrum is
\begin{align}
\mathcal{S}(\mathbf{K},\Omega)= {}&{}\int{d}t\,{d}^2\mathbf{r}\,e^{i\Omega{t}-i\mathbf{K}\mathbf{r}}
\times{}\nonumber\\
{}&{}\times
\left[\frac12{\langle \{\hat{J}(\mathbf{r},t),\hat{J}(0,0)\}\rangle}-\langle J \rangle^2\right].
\label{eq:2d_spec_def}
\end{align}
The details of the calculation are given in Appendix~\ref{app:layers}, here we sketch the main steps. Following the standard procedure found, for example, in Ref.~\cite{Kamenev2011}, we construct the generating functional for electrons including source fields that couple to electron density and current. Intra- and inter-layer Coulomb interactions are handled via the introduction of a Hubbard-Stratonovich field, which results in a nonlinear bosonic theory. Treating the interaction in the random phase approximation (RPA), we arrive at a quadratic bosonic theory, from which one can determine arbitrary moments of currents and densities via functional differentiation. Focusing once again on the case where $T_2\gg T_1$, we obtain the noise spectrum with a structure very similar to that of Eq.~\eqref{eq:spectrum_integral}:
\begin{align}
\mathcal{S}(\mathbf{K},\Omega)={}&{}
\int\frac{d\omega}{2\pi}\,\frac{d^2\mathbf{k}}{(2\pi)^2}
\left[\coth\frac{\hbar\omega}{2T_2}\coth\frac{\hbar(\Omega-\omega)}{2T_2}+1\right]
\nonumber\\ {}&{}\times
\hbar^2\omega(\Omega-\omega)\Re\sigma(\omega)\Re\sigma(\Omega-\omega)
\nonumber\\ {}&{}\times
\left[|\sigma(\omega)|^2+\sigma(\Omega-\omega)\,\sigma^*(\omega)\right]
\frac{|\mathbf{k}\cdot(\mathbf{K}-\mathbf{k})|^2}{k^2|\mathbf{K}-\mathbf{k}|^2}
 \nonumber \\ {}&{}\times
 \left| u_{12}(\mathbf{k},\omega)\right|^2 \left|u_{12}(\mathbf{K}-\mathbf{k},\Omega-\omega) \right|^2 .\label{eq:SKOmega=}
\end{align}
Again, for temperatures $\hbar/\tau\ll{T}_2\ll\hbar{v}_-/d$ when the antisymmetric surface plasmon dominates the heat transfer, the integrand may be approximated as two overlapping Lorentzians originating from the complex poles of $u_{12}(\mathbf{k},\omega)$ and located near the corresponding dispersion relation. Performing the integration for $\Omega\ll T_2/\hbar$ (see Appendix~\ref{app:layers} for details), we find
\begin{equation}
\mathcal{S}(\mathbf{K}, \Omega) = \frac{3 \zeta(3)\,T_2^4}{16\pi^2\hbar^2 \sigma_\mathrm{dc}d} \,\mathrm{Im} \left[\frac{1+(\Omega\tau/2)^2}{\sqrt{(\Omega\tau -i)^2 - (v_-K\tau)^2}}\right],
\label{eq:2d_res}
\end{equation}
our second main result.
The spectrum \eqref{eq:2d_res} is plotted in Fig.~\ref{fig:2d_res}; it exhibits a sharp feature near the plasmon dispersion relation $\Omega = v_- K$, whose width determined by the plasmon lifetime $\tau$. 
The shape of the feature is different from the circuit result~\eqref{eq:0d_res} due to the 2D geometry of the system.
At $K,\Omega\to0$, Eq.~\eqref{eq:2d_res} once again gives the heat current shot noise, $\mathcal{S}(0,0) = (3T_2/2) \langle J \rangle$; there is no parametrically large Fano factor since the heat is transferred by plasmons whose frequency $\omega\sim{T}_2/\hbar$.

\section{Observability of the heat current noise}
In experiments, the heat current is typically inferred from a measured change in the electronic temperature $T_1$ of the cold body.
One can speak about temperature fluctuations only at time scales exceeding the relaxation time of the electronic distribution function, which is always slower than the thermal time $\hbar/T_1$.
In addition to this fundamental requirement, the experimental observability of $\mathcal{S}(\mathbf{K},\Omega)$ is restricted more strongly by the relaxation dynamics of the temperature itself, which is governed by exchange of heat between electrons and phonons and by electronic thermal conductivity.

Let us first focus on the example of two-dimensional metallic layers, assuming that the phonons in the cold metal layer are in thermal equilibrium with the substrate/cryostat held at constant uniform temperature ${T}_\mathrm{ph}$. The electrons on the other hand are perturbed by the incoming radiation so they are not in thermal equilibrium with the phonons. The local electronic temperature $T_1(\mathbf{r}, t)$ satisfies the energy balance equation:
\begin{equation}
\mathcal{C}\, \frac{\partial T_1}{\partial t} = \kappa \nabla^2T_1 - J_\mathrm{ph} + J,
\label{eq:energy_balance}
\end{equation}
where $\mathcal{C}$ and $\kappa$ are electronic specific heat and thermal conductivity, respectively. $J_\mathrm{ph}$~is the power per unit area given by the electrons to the phonons; assuming small temperature change, we linearize $J_\mathrm{ph}=G_\mathrm{th}(T_1-{T}_\mathrm{ph}) + \delta J_\mathrm{ph}(\mathbf{r}, t)$ with $G_\mathrm{th}$ being the electron-phonon thermal conductance per unit area, and the fluctuating part $\delta J_\mathrm{ph}(\mathbf{r}, t)$ which satisfies the fluctuation-dissipation theorem $\langle\delta J_\mathrm{ph}(\mathbf{r}, t)\,\delta J_\mathrm{ph}(\mathbf{r}', t')\rangle = 2 T_\mathrm{ph}^2 G_\mathrm{th}\,\delta(\mathbf{r}-\mathbf{r}')\,\delta(t-t')$ valid up to frequencies $\Omega \lesssim T_\mathrm{ph}$~\cite{Pekola2018}.
$J(\mathbf{r}, t)$ is the fluctuating radiative heat current, here treated as a classical random variable with the average $\langle{J}\rangle$ and fluctuations $\delta{J}(\mathbf{r},t)$ whose spectrum $\mathcal{S}(\mathbf{K},\Omega)$ was calculated above. 
Then Eq.~(\ref{eq:energy_balance}) gives $T_1(\mathbf{r},t)=T_\mathrm{ph}+\langle{J}\rangle/G_\mathrm{th}+\delta{T}_1(\mathbf{r},t)$, where the temperature fluctuation spectrum $\mathcal{S}_T(\mathbf{K},\Omega)$, defined as the Fourier transform of $\langle\delta{T}_1(\mathbf{r}, t)\,\delta{T}_1(0,0)\rangle$ similarly to Eq.~(\ref{eq:2d_spec_def}), is given by
\begin{equation}
\mathcal{S}_{T}(\mathbf{K}, \Omega) = \frac{\mathcal{S}(\mathbf{K}, \Omega) + 2 T_\mathrm{ph}^2 G_\mathrm{th}}{\left(G_\mathrm{th}+\kappa K^2\right)^2 + \Omega^2 \mathcal{C}^2}.
\label{eq:T_spec}
\end{equation}
For $\mathcal{S}(\mathbf{K}, \Omega)$ to be measurable, (i)~it should not be overwhelmed by the phonon contribution in the numerator (this is where having $T_2\gg{T}_1\approx{T}_\mathrm{ph}$ is useful), and (ii)~the denominator should be roughly flat on the scales $K\sim 1/(v_-\tau)$ and $\Omega\sim 1/\tau$, corresponding to the interesting features of $\mathcal{S}(\mathbf{K},\Omega)$, 
so we need $1/\tau \lesssim G_\mathrm{th}/\mathcal{C}$ and $1/(v_-\tau) \lesssim \sqrt{G_\mathrm{th}/\kappa}$.

For an estimate, we consider doped graphene monolayers where plasmon dominated radiative heat transfer has been observed (see the discussion in Sec. 2.3 of Ref.~\cite{Yang2018}). Taking the Fermi energy $E_F = 100\:\mbox{meV}$ and the electron mean free path $v_F\tau = 3\:\mu\mbox{m}$ with $v_F=10^8\:\mbox{m/s}$, we estimate the thermal conductivity via the Wiedemann-Franz law as $\kappa = (\pi/3)E_F\tau{T}_1 /\hbar^2$,  and the heat capacity per unit area is found for 2D Dirac electrons as $\mathcal{C} = (2\pi/3)E_F T_1 /(\hbar{v}_F)^2$. 
$G_\mathrm{th}$ may be estimated from Ref.~\cite{Fong2013} reporting $J_\mathrm{ph} = \Sigma(T_1^3 - T_\mathrm{ph}^3)$ with $\Sigma \approx 1.25 \mbox{W} \mbox{K}^{-3} \mbox{m}^{-2}$. This gives the frequency scale in the denominator of Eq.~(\ref{eq:T_spec}) $G_\mathrm{th}/\mathcal{C} = 6.5 \times 10^9\,[T_1/(1\:\mbox{K})]\:\mbox{s}^{-1}$. Then condition (ii) on frequency $1/\tau \lesssim G_\mathrm{th}/\mathcal{C}$ is satisfied for temperatures $T_1 \gtrsim 50\:\mbox{K}$. In this regime, the corresponding condition on wavevector $1/(v_-\tau) \lesssim \sqrt{G_\mathrm{th}/\kappa}$ is satisfied automatically since for $G_\mathrm{th}/\mathcal{C} \sim 1/\tau$, we have $\sqrt{G_\mathrm{th}/\kappa} \sim 1/(v_F \tau) \gg 1/(v_- \tau)$. 
Condition (i) regarding the numerator of Eq.~(\ref{eq:T_spec}) is easily satisfied -- for example, taking $T_2 = 300\:\mbox{K}$ and $d = 430\:\mbox{nm}$ as in Ref.~\cite{Yang2018} we find $\mathcal{S}(0,0)/(T_\mathrm{ph}^2G_\mathrm{th}) \sim 10^4$.

For a zero-dimensional quantum circuit, we use Eq.~\eqref{eq:T_spec} with $K=0$ and replace $\mathcal{S}(\mathbf{K},\Omega)$ in the numerator by $\mathcal{S}(\Omega)$ from Eq.~\eqref{eq:0d_res}. Condition~(ii) becomes $\gamma \ll G_\mathrm{th}/\mathcal{C}$ allowing much freedom: indeed, $G_\mathrm{th}/\mathcal{C}$ is an intrinsic material property of the resistors only ($G_\mathrm{th}$ and $\mathcal{C}$ are both proportional to volume), while $\gamma$ depends only on the resonator and its coupling to the resistors, which is often tunable \cite{Meschke2006}.  Taking material parameters for copper from Ref.~\cite{Karimi2020}, we find the denominator frequency scale $G_\mathrm{th}/\mathcal{C} = 6\times 10^8\,[T_1/(1\:\mbox{K})]^3\:\mbox{s}^{-1}$. For setups like that of Ref.~\cite{Ronzani2018} with resonator frequency in the $\mbox{GHz}$ range and quality factors $10-100$, condition~(ii) can be satisfied at the relevant sub-Kelvin temperatures. Condition~(i) can be satisfied by choosing sufficiently different $T_1$ and $T_2$; e.~g., for copper the two contributions in the numerator of Eq.~(\ref{eq:T_spec}) are comparable for $T_1 = 0.1\:\mbox{K}$ and $T_2 = 1\:\mbox{K}$.


\section{Conclusions}
We have shown how noise spectrum of the radiative heat curent may shed light on the nature of heat-carrying excitations. Namely, when the heat is carried by a single mode of a resonator, or by a well-defined branch of surface excitations, the noise spectrum contains a peak at low frequencies, whose width is related to the excitation lifetime.
We found that observability of such feature via temperature measurements is determined by a trade-off in terms of the thermal conductance between electrons and phonons,~$G_\mathrm{th}$. For too large $G_\mathrm{th}$, the electron temperature fluctuations are dominated by the electron-phonon heat exchange which mask the radiative contribution. On the other hand, if $G_\mathrm{th}$ is too small, the resonant feature in the radiative heat current noise spectrum is not resolved because the thermal response of the metal is too slow, so the heat current fluctuations are effectively averaged out. Such trade-off is possible for realistic parameters typical of doped graphene or of superconducting resonators.
An interesting possibility to measure the noise is to mimic the cold bath by a microwave transmission line, so the heat current can be accessed by measuring statistics of photons emitted into the transmission line~\cite{Gabelli2004}.

\acknowledgments 
We thank G. Catelani and J.~Pekola for insightful discussions and  C.~Altimiras for drawing our attention to Ref.~\cite{Gabelli2004}. This project received funding from the European Union’s Horizon 2020 research and innovation programme under the Marie Sk\l{o}dowska-Curie Grant Agreement No. 766025. WB acknowledges support by the Deutsche Forschungsgemeinschaft (DFG; German Research Foundation) via SFB 1432 (Project \mbox{No.} 425217212) and the EU’s Horizon 2020 research and innovation program under Grant Agreement No. 964398 (SUPERGATE).

\appendix

\section{Heat current noise in a quantum circuit}
\label{app:circuit}

\subsection{Quantum circuit model}
\label{sec:Quantum_circuit_model}
We first describe independently different elements of the circuit: the $LC$ resonator and the resistors.
The $LC$ loop has one degree of freedom which we chose to described by the charge $Q$ on the capacitor. The dynamics of this loop is governed by the Lagrangian:
\begin{equation}
    \mathcal{L}_{LC}=\frac{L}{2}\,\Dot{Q}^2-\frac{Q^2}{2C},
\end{equation}
where the kinetic term represents the energy stored in the inductance~$L$, and the potential term corresponds to the electrostatic energy of the capacitance~$C$. The $LC$ circuit realizes a harmonic oscillator which is quantized in the standard way: the conjugate variable to the charge $Q$ is the magnetic flux in the inductance, $\Phi=\partial\mathcal{L}_{LC}/\partial{\Dot{Q}}$, and the canonical commutation relation is imposed: $[\hat{Q},\hat{\Phi}]=i\hbar$. Introducing the resonator photon creation and annihilation operators $\hat{a}_{LC}^\dagger,\hat{a}_{LC}$, as
\begin{equation}
\hat{a}_{LC}=\frac{(L/C)^{1/4}}{\sqrt{2\hbar}}\,\hat{Q}+\frac{(C/L)^{1/4}}{\sqrt{2\hbar}}\,i\hat{\Phi},
\end{equation}
we deduce the Hamiltonian operator as a function of $Q$ and $\Phi$, or of $\hat{a}_{LC}^\dagger,\hat{a}_{LC}$:
\begin{equation}
    \hat{\mathcal{H}}_{LC}=\frac{\hat{\Phi}^2}{2L}+\frac{\hat{Q}^2}{2C}=\hbar\omega_0\left(\hat{a}_{LC}^\dagger \hat{a}_{LC}+\frac{1}{2}\right), \quad\omega_0\equiv\frac1{\sqrt{LC}}.
\end{equation}

\begin{figure}
\includegraphics[width = 0.44\textwidth]{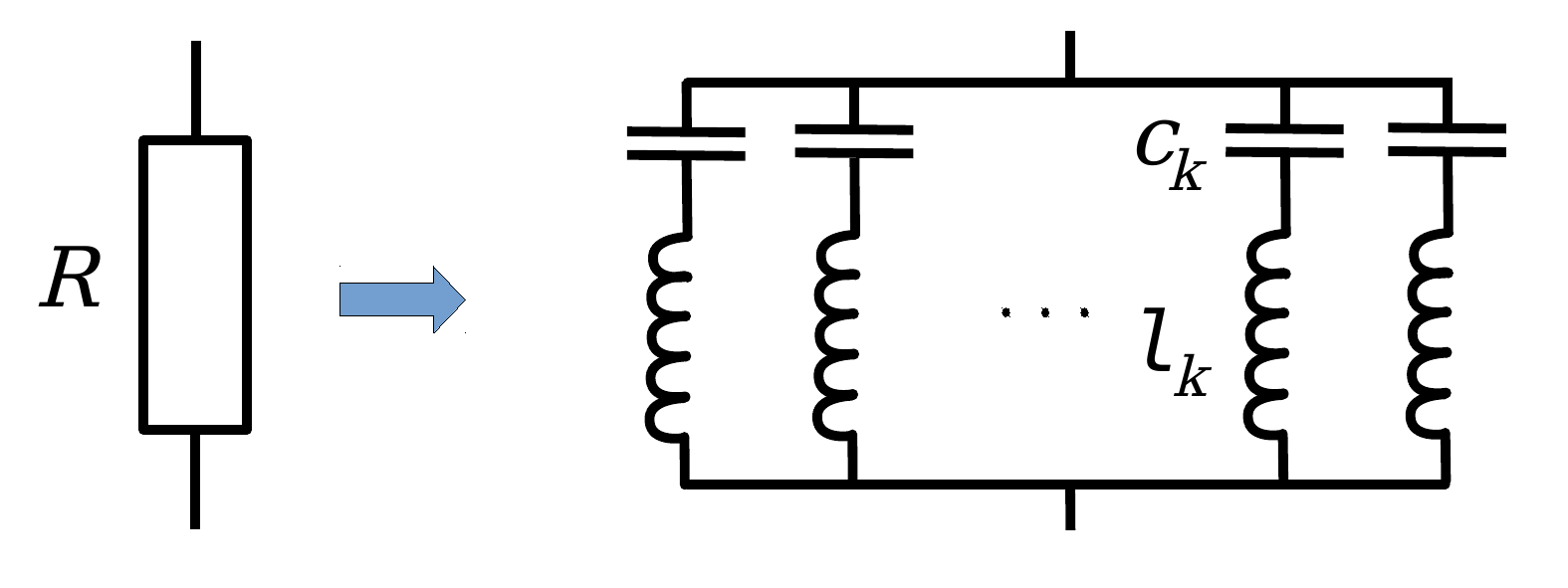}
\caption{A resistor modelled as a large collection of $LC$ oscillators.}
\label{fig:Resistor}
\end{figure}

Each resistor represents a heat bath kept at a fixed temperature. Each bath has a huge number of degrees of freedom which can be modeled as independent harmonic oscillators. 
In this way, the Lagrangian and the Hamiltonian of a resistor~$R$ can be written as
\begin{align}
    \mathcal{L}_R&=\sum_{k=1}^{N} \left(\frac{l_k}{2}\Dot{q}_k^2-\frac{q_k^2}{2c_k}\right), \\
    \hat{\mathcal{H}}_R&=\sum_{k=1}^{N}\left(\frac{\hat{\phi}_k^2}{2l_k}+\frac{\hat{q}_k^2}{2c_k}\right)=\sum_{k=1}^{N} \hbar\omega_k\left(\hat{a}_k^\dagger\hat{a}_k+\frac{1}{2}\right),
\end{align}
where $\omega_k=1/\sqrt{l_kc_k}$.
The coefficients $l_k$ and $c_k$ can be linked to the macroscopic resistance $R$ by requiring that the current through the short-circuited resistor,
\begin{equation}
\hat{I}=\sum_{k} \frac{d\hat{q}_k}{dt}=\sum_{k} \frac{i}\hbar\,[\hat{\mathcal{H}}_R,\hat{I}]=
-i\sum_{k}\sqrt{\frac{\hbar\omega_k}{2l_k}}\left(\hat{a}_k-\hat{a}_k^\dagger\right),
\end{equation}
satisfies the fluctuation-dissipation theorem:
\begin{align}
    \frac{1}{2}\left\langle \{\hat{I}(t),\hat{I}(0)\} \right\rangle={}&{}
    \sum_{k=1}^N\frac{\hbar\omega_k}{l_k}\frac{\cos\omega_kt}{2}\,\coth{\frac{\hbar\omega_k}{2T}}
    ={}\nonumber\\
    ={}&{}\int_{-\infty}^{+\infty}\frac{d\omega}{2\pi}\,e^{-i\omega{t}}\,
    \frac{\hbar\omega}{R}\coth\frac{\hbar\omega}{2T}.
\label{fluct_diss_theorem_current}
\end{align}
Here the first expression is obtained using $\hat{a}_k(t)=\hat{a}_ke^{-i\omega_kt}$ and the standard relations for harmonic oscillators averages over the thermal state:
\begin{subequations}\label{eqs:thermalbath}
\begin{align}
&    \langle \hat{a}_{k} \rangle =\langle \hat{a}_{k}^\dagger \rangle =0,\quad
    \langle \hat{a}_{k} \hat{a}_{k'} \rangle=\langle \hat{a}_{k}^\dagger \hat{a}_{k'}^\dagger \rangle=0, \\
&    \langle \hat{a}_{k}^\dagger \hat{a}_{k'} \rangle=\frac{\delta_{kk'}}{e^{\hbar\omega_{k}/T}-1}, \quad
    \langle \hat{a}_{k} \hat{a}_{k'}^\dagger \rangle=\frac{\delta_{kk'}}{1-e^{-\hbar\omega_{k}/T}}. \label{a_correlation}
\end{align}
\end{subequations}
Identifying the two expressions in Eq.~(\ref{fluct_diss_theorem_current}), we obtain
\begin{equation}\label{eq:Resistance}
    \frac{1}{R}=\lim_{\eta\to 0^+}\lim_{N \to \infty}\sum_{k=1}^N\frac{1}{l_k}\frac{i\omega}{\omega^2-\omega_k^2+i\eta\omega},
\end{equation}
where the positive infinitesimal $\eta$ is needed to ensure the smooth limit $N\to\infty$.
Noting that Eq.~(\ref{eq:Resistance}) can also be written as $1/R=\sum_k[-i\omega{l}_k-1/(i\omega{c}_k)]^{-1}$, we can imagine the resistor as a parallel array of inductors and capacitors, shown in Fig~\ref{fig:Resistor}.

The total Lagrangian of our circuit is obtained by adding the coupling via mutual inductances $M_1$ and $M_2$ to the independent Lagrangians of the three subsystems:
\begin{equation}
    \mathcal{L}=\mathcal{L}_{LC}+\mathcal{L}_{R_1}+\mathcal{L}_{R_2}+M_1\Dot{Q}\sum_{k=1}^{N}\Dot{q}_{1,k}+M_2\Dot{Q}\sum_{k=1}^{N}\Dot{q}_{2,k}.
\end{equation}
In fact, the sign of the coupling terms is arbitrary, since it depends on the arbitrary choice of the current direction in the three loops.
This full Lagragian is quadratic, so it can be written in the matrix form:
\begin{equation}
\mathcal{L}=\frac{1}{2}\sum_{i,j}\left(L_{ij}\Dot{q}_i\Dot{q}_j-C^{-1}_{ij}q_iq_j\right)
\equiv\frac{1}2\left(\underline{\dot{q}}^T\underline{\underline{L}}\,\underline{\dot{q}}
-\underline{q}^T\underline{\underline{C}}^{-1}\,\underline{q}\right),
\end{equation}
where $\underline{q}=(\ldots q_{1,k}  \ldots Q \ldots q_{2,k}\ldots)^T$, the capacitance matrix 
$\underline{\underline{C}}=\diag(\hdots c_{1,k}  \hdots  C  \hdots c_{2,k}  \hdots)$, and the inductance matrix
\begin{align}
    \underline{\underline{L}}=\begin{pmatrix}
    \ddots & 0 & {} & M_1 & {} & {} & {} \\
    0 & l_{1,k} & {} & \vdots & {} & 0 & {} \\
    {} & {} & \ddots & M_1 & {} & {} & {} \\
    M_1 & \hdots & M_1 & L & M_2 & \hdots & M_2 \\
    {} & {} & {} & M_2 & \ddots & 0 & {} \\ 
    {} & 0 & {}& \vdots & 0 & l_{2,k} & {} \\ 
    {} & {} & {} & M_2 & {} & {} & \ddots\\
    \end{pmatrix}.
\end{align}
From the full Lagrangian, we deduce the new set of conjugate variables of the entire circuit: $\phi_i=\partial\mathcal{L}/\partial{\Dot{q}_i}=\sum_j L_{ij}\Dot{q}_j$, and finally the total quantum Hamiltonian:
\begin{align}
    \mathcal{\hat{H}}={}&{}\frac{1}{2}\sum_{i,j}\left(L^{-1}_{ij}\hat{\phi}_i\hat{\phi}_j+C^{-1}_{ij}\hat{q}_i\hat{q}_j\right)\equiv\nonumber\\
    \equiv{}&{}\frac{1}2\left(\underline{\phi}^T\underline{\underline{L}}^{-1}\underline{\phi}
-\underline{q}^T\underline{\underline{C}}^{-1}\,\underline{q}\right).
\label{eq:circuitHamiltonian}
\end{align}
Note that unlike $\underline{\underline{L}}$, the inverse $\underline{\underline{L}}^{-1}$ is a full matrix. As a result, all $\hat\phi_i$ variables are coupled to each other, so the coupling between the three subsystems appears to be nonlocal. This is because the $\hat\phi_i$ variables are themselves nonlocal (that is, they cannot be ascribed to just one of the three subsystems), and the coupling remains local when expressed in terms of the local variables $d\hat{q}_i/dt$.

\subsection{Quantum Langevin equations}

Hamiltonian (\ref{eq:circuitHamiltonian}) gives the following Heisenberg's equations of motion:
\begin{subequations}\begin{align}
    \frac{d\hat{q}_i(t)}{dt}&=\frac{i}{\hbar}[\mathcal{\hat{H}},\hat{q}_i(t)]=\sum_j L_{ij}^{-1}\hat{\phi}_j(t), \\
    \frac{d\hat{\phi}_i(t)}{dt}&=\frac{i}{\hbar}[\mathcal{\hat{H}},\hat{\phi}_i(t)]=-\sum_j C_{ij}^{-1}\hat{q}_j(t),
\end{align}\end{subequations}
for all $i$. Elimination of the $\hat{\phi}_i$'s yields ($\lambda=1,2$ labels the resistors):
\begin{subequations}\label{eqs:d2qdt2}\begin{align}
&    l_{\lambda,k}\,\frac{d^2\hat{q}_{\lambda,k}(t)}{dt^2}+\frac{\hat{q}_{\lambda,k}(t)}{c_{\lambda,k}}=-M_\lambda\,\frac{d^2\hat{Q}(t)}{dt^2},\label{Heisenberg_charges}\\
& L\,\frac{d^2\hat{Q}(t)}{dt^2}+\frac{\hat{Q}(t)}{C}=-\sum_{\lambda=1,2}\sum_{k=1}^{N}M_\lambda\,\frac{d^2\hat{q}_{\lambda,k}(t)}{dt^2}.
\end{align}\end{subequations}
Next, we invert the differential operators on the left-hand side of these equations, and pass to the integral form. Namely, treating the right-hand side of each equation as an external force, we assume each solution to be the sum of a free oscillation and of the one induced by the external force.
This procedure is done more conveniently introducing Fourier components,
\begin{equation}
\hat{q}_i(\omega)=\int\hat{q}_i(t)\,e^{i\omega{t}}\,dt,
\end{equation}
so that Eqs.~(\ref{eqs:d2qdt2}) become
\begin{subequations}\begin{align}
 &   \hat{q}_{\lambda,k}(\omega)=\hat{q}_{1,k}^{(0)}(\omega)-\frac{M_\lambda}{l_{1,k}}\frac{\omega^2\hat{Q}(\omega)}{(\omega+i0^+)^2-\omega_{\lambda,k}^2}, \label{q1} \\
  &  \hat{Q}(\omega)=-\frac{\omega^2}{(\omega+i0^+)^2-\omega_0^2}\sum_{\lambda=1,2}\sum_{k=1}^{N}\frac{M_\lambda}{L}\,\hat{q}_{\lambda,k}(\omega). \label{Q}
 \end{align}\end{subequations}
The infinitesimal imaginary part in the denominator ensures causality (the induced oscillation is determined by the force in the past, but not in the future).
The free oscillation part is given by
\begin{align}
\hat{q}_{\lambda,k}^{(0)}(\omega)={}&{}\sqrt{\frac{\hbar}{2\omega_{\lambda,k}l_{\lambda,k}}}
\times{}\nonumber\\
{}&{}\times\left[2\pi\delta(\omega-\omega_{\lambda,k})\hat{a}_{\lambda,k}+2\pi\delta(\omega+\omega_{\lambda,k})\hat{a}_{\lambda,k}^\dagger\right], \label{Langevin} 
\end{align}
where $\hat{a}_{\lambda,k},\hat{a}_{\lambda,k}^\dagger$ are taken to be the operators of the non-interacting thermal baths, satisfying Eqs.~(\ref{eqs:thermalbath}) for each bath~$\lambda$ with temperature~$T_\lambda$. This corresponds to the assumption that the baths have so large number of degrees of freedom, that their initial thermal equilibrium remains unaffected, even long time after the coupling was switched on in the remote past. At the same time, for the $LC$ oscillator we include no free oscillation, assuming that its initail condition is forgotten since the coupling was switched on.

The free oscillations of the bath give rise to the Langevin forces in the Kirchhoff's laws for the circuit currents $\hat{I}_{LC}(\omega)=-i\omega\hat{Q}(\omega)$ (the current in the $LC$~loop) and $\hat{I}_\lambda(\omega)=-i\omega\sum_k\hat{q}_{\lambda,k}(\omega)$ (the current through the reistor~$\lambda$).  Let us sum over $k$ in Eq.~(\ref{q1}) and recognize Eq.~(\ref{eq:Resistance}) for $R_1$ and $R_2$:
\begin{subequations}\label{eqs:Kirchhoff}\begin{align}
&   \hat{I}_\lambda(\omega)=\frac{i\omega{M}_\lambda}{R_\lambda}\,\hat{I}_{LC}(\omega)+\hat{I}_\lambda^{(0)}(\omega),  \label{Kirchhoff2}\\
   & \left(i\omega L+\frac{1}{i\omega C}\right)\hat{I}_{LC}(\omega)=-i\omega M_1\hat{I}_1(\omega)-i\omega M_2\hat{I}_2(\omega),
\end{align}\end{subequations}
where $\hat{I}^{(0)}_\lambda(\omega)=-i\omega\sum_k\hat{q}^{(0)}_{\lambda,k}$ are the Langevin forces (fluctuating current sources). Their average $\langle \hat{I}_\lambda^{(0)}(t) \rangle =0$, and their pair correlators are straighforwardly evaluated from Eqs.~(\ref{eqs:thermalbath}):
\begin{subequations}\label{eq:II}\begin{align}
&\left\langle\left\{\hat{I}_\lambda^{(0)}(t),\hat{I}_{\lambda'}^{(0)}(0)\right\}\right\rangle=
\int^{\infty}_{-\infty}\frac{d\omega}{2\pi}\,i\hbar\Pi_{\lambda\lambda'}^K(\omega)\,e^{-i\omega{t}},\\
&\left\langle\left[\hat{I}_\lambda^{(0)}(t),\hat{I}_{\lambda'}^{(0)}(0)\right]\right\rangle=
\int^{\infty}_{-\infty}\frac{d\omega}{2\pi}\,i\hbar\Pi_{\lambda\lambda'}^{RA}(\omega)\,e^{-i\omega{t}},\end{align}\begin{align}
&\Pi_{\lambda\lambda'}^K(\omega)=-2i\delta_{\lambda\lambda'}\frac\omega{R_\lambda}\coth{\frac{\hbar\omega}{2T_\lambda}},\\
&\Pi_{\lambda\lambda'}^{RA}(\omega)=-2i\delta_{\lambda\lambda'}\frac\omega{R_\lambda},\label{eq:PiKRA}
\end{align}\end{subequations}
so that the Langevin forces are indeed independent and satisfy the fluctuation-dissipation theorem.
The superscripts ``$K$'' and ``$RA$'' here are just notations; they are chosen to match those of the non-equilbrium Green's functions, used in Appendix~\ref{app:layers}. $\Pi_{\lambda\lambda'}^K(\omega)$ introduced in this way represents the Keldysh component of the current-current polarisation operator, while $\Pi_{\lambda\lambda'}^{RA}(\omega)$ is the difference between its retarded and advanced components. Higher-order moments of the Langevin forces are reduced to products of these pair averages using the Wick's theorem, since the forces are linear combinations of free bosonic operators and the averaging is performed over the thermal state.

\subsection{Operator of the dissipated power}

The observable we are interested in is the power dissipated in resistor 1: 
\begin{gather}
    \hat{P}=\frac{d}{dt}\sum_{k=1}^N\left[\frac{l_{1,k}}{2}\left(\frac{d\hat{q}_{1,k}}{dt}\right)^2 +\frac{\hat{q}_{1,k}^2}{2c_{1,k}}\right],
\end{gather}
where the time derivatives are determined by the commutator with the total Hamiltonian~$\hat{\mathcal{H}}$. The sum represents the energy stored in the resistor's internal degrees of freedom; we write it in terms of the charge variable without involving the fluxes, since the latter are nonlocal, as discussed in Sec.~\ref{sec:Quantum_circuit_model}.
Applying the commutators and using the equations of motion (\ref{Heisenberg_charges}), we obtain
Eq.~(\ref{eq:Pcircuit}) of the main text.
This equation expresses the fact that the heat flux flowing from $R_2$ to $R_1$ is given by the work performed on the current $\hat{I}_{1}$ by the external voltage induced by the mutual inductance, $-M_1d\hat{I}_{LC}/dt$. The corresponding quantum operator is of course Hermitian, which is ensured by its anticommutative form. 


Given this form of the power operator, it is convenient to solve the linear equations~(\ref{eqs:Kirchhoff}) for $\hat{I}_{1,2}(\omega)$ and $\hat{I}_{LC}(\omega)$, and express the two operators $\hat{I}_1$ and $-M_1d\hat{I}_{LC}/dt$ in terms of $\hat{I}^{(0)}_\lambda(\omega)$. We introduce the matrix notation for the solutions:
\begin{subequations}\begin{align}
&i\omega{M}_\lambda\hat{I}_{LC}(\omega)=\sum_{\lambda'=1,2}U_{\lambda\lambda'}(\omega)\,\hat{I}^{(0)}_{\lambda'}(\omega),\\
&\hat{I}_\lambda(\omega)=\sum_{\lambda'=1,2}\Xi_{\lambda\lambda'}(\omega)\,\hat{I}^{(0)}_{\lambda'}(\omega),
\end{align}\end{subequations}
where we defined two matrices $U(\omega)$ and $\Xi(\omega)$ (it is the off-diagonal matrix element $U_{12}(\omega)$ that is given by Eq.~(\ref{eq:U_12}) of the main text):
\begin{subequations}\label{eqs:Umatrix}\begin{align}
& U(\omega)
=\frac{i\omega^3C}{\Delta(\omega)}
\begin{pmatrix} M_1^2 & M_1M_2 \\ M_1M_2 & M_2^2 \end{pmatrix},\\
& \Xi(\omega)
=\begin{pmatrix} 1 & 0 \\ 0 & 1 \end{pmatrix}+
\begin{pmatrix} 1/R_1 & 0 \\ 0 & 1/R_2 \end{pmatrix}U(\omega),\\
&\Delta(\omega)\equiv1-\omega^2LC-i\omega^3C(M_1^2/R_1+M_2^2/R_2).
\end{align}\end{subequations}

\subsection{Average transferred power}

The average power in terms of the correlation functions follows straightgorwardly:
\begin{align}
    \langle \hat{P} \rangle={}&{}\int_{-\infty}^{\infty}\frac{d\omega}{2\pi}\,
    \sum_{\lambda,\lambda'=1,2}\Xi_{1\lambda}(\omega)\,\frac{i\hbar}2\,\Pi^K_{\lambda\lambda'}(\omega)\,U_{1\lambda'}^*(\omega)={}\nonumber\\
    ={}&{}\int_{-\infty}^{\infty}\frac{d\omega}{2\pi}\,|U_{12}(\omega)|^2\,\frac{\hbar\omega}{R_1R_2}\left(\coth\frac{\hbar\omega}{2T_2}-\coth\frac{\hbar\omega}{2T_1}\right).
\end{align}
The Landauer form of this equation is given by Eq.~(\ref{eq:Landauer}) of the main text. $U_{12}(\omega)$~is a rational function while the difference of the hyperbolic cotangents decreases exponentially for $\omega>\max\{T_1,T_2\}/\hbar$. Let us rewrite identically
\begin{equation}
U_{12}(\omega)=\frac{i\omega^3\sqrt{\gamma_1\gamma_2}\sqrt{R_1R_2}}{\omega_0^2(\omega_0^2-\omega^2)-i\omega^3(\gamma_1+\gamma_2)}.
\end{equation}
When the coupling is weak ($\gamma\ll\omega_0$) we can distinguish three regimes where different terms dominate in the denominator of $U_{12}(\omega)$.
\begin{itemize}
\item
In the low-temperature regime, $T_{1,2}\ll\hbar\omega_0$, the denominator can be approximated simply by~$\omega_0^4$, so  it is the low-frequency behaviour of $|U_{12}(\omega)|^2\propto\omega^6$ that is important, which gives
    \begin{equation}
        \langle P \rangle_{\text{lt}} = \frac{16{\hbar\gamma_1\gamma_2}}{15\pi}\,\frac{T_2^8-T_1^8}{(\hbar\omega_0/\pi)^8}.
    \end{equation}
\item
In the intermediate-temperature regime $\hbar\omega_0\ll T_{1,2}\ll\hbar\omega_0^2/\gamma$, the denominator is dominated by $\omega_0^2\omega^2$, which gives
    \begin{equation}\label{P:high-t-regime}
        \langle P \rangle_{\text{it}} = \frac{2\pi{\hbar\gamma_1\gamma_2}}{15}\,\frac{T_2^4-T_1^4}{(\hbar\omega_0)^4}.
    \end{equation}
\item 
In the high-temperature regime $T_{1,2}\gg \hbar\omega_0^2/\gamma$, where the denominator is dominated by the last term $\propto\omega^3$:
    \begin{equation}\label{P:very-high-t-regime}
        \langle P \rangle_{\text{ht}}= \frac{\pi}{3}\frac{\gamma_1\gamma_2}{\hbar\gamma^2}\,(T_2^2-T_1^2).
    \end{equation}
\end{itemize}
In both the intermediate and the high temperature regimes, one should add the contribution from the sharp Lorentzian resonance at $\omega=\omega_0$ [we approximate $\coth[\hbar\omega/(2T)]\approx2T/(\hbar\omega_0)$]:
\begin{align}
\langle P\rangle_{\text{res}}=\int_0^{\infty}\frac{d\omega}{\pi}\,\frac{2(T_2-T_1)\omega_0^2\gamma_1\gamma_2}{(\omega^2-\omega_0^2)^2+\omega_0^2\gamma^2}=\frac{\gamma_1\gamma_2}{\gamma}\,(T_2-T_1).
\end{align}
Since the resonance width $\gamma$ does not appear in the denominator of (\ref{P:high-t-regime}) but does in (\ref{P:very-high-t-regime}), the resonant contribution only dominates in the intermediate-temperature regime, when $T_{1,2}\ll\hbar\omega_0(\omega_0/\gamma)^{1/3}$.

In cases where the two temperatures do not belong to the same regime, it is the hottest resistor that  determines the regime. For example, under the conditions $T_1 \ll \hbar\omega_0 \ll T_2 \ll \hbar\omega_0(\omega_0/\gamma)^{1/3}$, we have $\langle P \rangle=(\gamma_1\gamma_2/\gamma)T_2$.

\subsection{Transferred power noise spectrum}

The noise correlator [Eq.~(\ref{eq:0d_spec_def}) of the main text] involves averages of four operators which we express in terms of the pair averages using the Wick's theorem: $\langle \hat{x}_1\hat{x}_2\hat{x}_3\hat{x}_4 \rangle=\langle \hat{x}_1\hat{x}_2 \rangle \langle \hat{x}_3\hat{x}_4 \rangle+\langle \hat{x}_1\hat{x}_3 \rangle\langle \hat{x}_2\hat{x}_4 \rangle+\langle \hat{x}_1\hat{x}_4 \rangle\langle \hat{x}_2\hat{x}_3 \rangle$. Then, using Eqs.~(\ref{eq:II}) and rearranging the terms, we obtain
\begin{align}
\mathcal{S}(\Omega)={}&-\frac{\hbar^2}4\int\limits_{-\infty}^\infty \frac{d\omega}{2\pi} 
\times{}\nonumber\\ &\times
\Big[(\Xi\Pi^K\Xi^\dagger)_{11}(\omega)\,(U\Pi^KU^\dagger)_{11}(\Omega-\omega)
+{}\nonumber\\
&\quad{}+(\Xi\Pi^KU^\dagger)_{11}(\omega)\,(U\Pi^K\Xi^\dagger)_{11}(\Omega-\omega)
+{}\nonumber\\
&\quad{}+(\Xi\Pi^{RA}\Xi^\dagger)_{11}(\omega)\,(U\Pi^{RA}U^\dagger)_{11}(\Omega-\omega)
+{}\nonumber\\
&\quad{}+(\Xi\Pi^{RA}U^\dagger)_{11}(\omega)\,(U\Pi^{RA}\Xi^\dagger)_{11}(\Omega-\omega)\Big].
\label{eq:XiPiXi}
\end{align}
Here each $\Pi^K$ or $\Pi^{RA}$ is a diagonal matrix, so the integrand can be grouped in four terms, proportional to
\begin{align*}
&\coth\frac{\hbar\omega}{2T_1}\coth\frac{\hbar(\Omega-\omega)}{2T_1}+1,\\
&\coth\frac{\hbar\omega}{2T_1}\coth\frac{\hbar(\Omega-\omega)}{2T_2}+1,\\
&\coth\frac{\hbar\omega}{2T_2}\coth\frac{\hbar(\Omega-\omega)}{2T_1}+1,\\
&\coth\frac{\hbar\omega}{2T_2}\coth\frac{\hbar(\Omega-\omega)}{2T_2}+1.
\end{align*}
We are interested in the case $T_2\gg{T}_1$, so the dominant contribution comes from the last line, and one should pick up terms involving $\Pi^K_{22}$ or $\Pi^{RA}_{22}$. Inserting the corresponding matrix elements of the matrices~(\ref{eqs:Umatrix}), we arrive at Eq.~(\ref{eq:spectrum_integral}) of the main text.

We note that while the $\coth\coth$ contribution comes from $\Pi^K$, the average of the anticommutator, the unity comes from $\Pi^{RA}$, the average of the commutator. The latter vanishes in the classical limit. At the same time, it is essential for the frequency integral to converge.
Thus, although the classical circuit laws and the assumption of the fluctuation-dissipation theorem are all we needed to determine the average value of the transferred power, the quantum formalism is necessary to correctly calculate its noise spetrum.

Our main interest is in the intermediate temperature regime, $\hbar\omega_0\ll T_{1,2}\ll\hbar\omega_0^2/\gamma$, when the average power is dominated by the resonant contribution. 
To evaluate the frequency integral, we simplify $|U_{12}(\omega)|^2|U_{12}(\Omega-\omega)|^2$ in the same way we did for the average in the previous section. The two factors have Lorentzian peaks at $\omega=\pm\omega_0$ and $\omega=\Omega\pm\omega_0$, so we have a resonant contribution coming from their overlap:
\begin{align}
S_{\text{res}}(\Omega)={}&{}\int_0^\infty\frac{d\omega}{\pi}\,
\frac{8T_2^2\gamma_1^2\gamma_2^2}{4(\omega-\omega_0)^2+\gamma^2}\times{}\nonumber\\
&\qquad\times\frac1{4(\omega-\Omega-\omega_0)^2+\gamma^2}={}\nonumber\\
={}&{}\frac{2\gamma}{\Omega^2 + \gamma^2}\,\left(\frac{\gamma_1 \gamma_2}{\gamma}\, T_2\right)^2.\label{peak_in_S}
\end{align}
The resonant contribution to the noise spectrum appears on top of a background, which can be calculated in the intermediate temperature regime by keeping the $\omega^2$ in the denominator of $U_{12}$. Then the integral is dominated by a wide range of frequencies, $\omega\sim{T}_2/\hbar$, so for $|\Omega|\ll{T}_2/\hbar$ the background contribution is effectively constant:
\begin{equation}
    S_{\text{it}}(\Omega)=\frac{128\pi^5}{21}\,\frac{\gamma_1^2\gamma_2^2}{\hbar^5\omega_0^8}\,T_2^7 .
\end{equation}
Its height is below that of the resonant peak (\ref{peak_in_S}) if $T_2\ll \hbar\omega_0(\omega_0/\gamma)^{3/5}$; still, even if this condition is not satisfied, the resonant peak may still be observable since a constant background can usually be subtracted quite efficiently.

\section{Heat current noise for two-dimensional metallic layers}
\label{app:layers}

In this appendix we detail some of the longer calculations required for the example of two-dimensional metallic layers exchanging heat via strongly coupled surface plasmons. We start with the derivation of the radiative heat current operator between two general electronic systems in the Coulomb limit. We then sketch the calculation of the generating functional for an interacting disordered electron system, defining the nonequilibrium (Keldysh) Green's function components and showing how they are screened due to interactions in the random phase approximation (RPA). Via functional differentiation of the generating functional, we calculate a general expression for the symmetrised correlator characterising the heat current fluctuations. Finally, specifying to the case of two-dimensional metallic layers, we calculate the heat current noise spectrum, revealing a resonant feature associated with the plasmonic character of the heat transfer.

\subsection{Heat current operator}
\label{Ann:heat_curr}

In this section we use the notation $\mathbf{r}=(x,y,z)$ for the three-dimensional position vector, denoting the position in the layer plane by $\mathbf{r}_\|=(x,y)$.
Here we derive the operator corresponding to the Joule losses in the Coulomb limit given in the main text by Eq.~\eqref{eqn:heatop}. We start by considering a general system of electrons in an  external potential which can be due to impurities or confinement, with non-relativistic Coulomb interaction. Such system is described by the following Hamiltonian:
\begin{subequations}
\begin{align}
\hat{H} = {}&{}\int d^3\mathbf{r}\, \hat\Psi^\dagger(\mathbf{r})\left[-\frac{\hbar^2\nabla^2}{2m} + e\varphi_\mathrm{ext}(\mathbf{r})\right]\hat\Psi(\mathbf{r}) + {}\nonumber\\
{}&{}+ \frac{e^2}{2}\int d^3\mathbf{r}\, d^3\mathbf{r}'\, 
\frac{\hat\Psi^\dagger(\mathbf{r})\,\hat\Psi^\dagger(\mathbf{r}')\,\hat\Psi(\mathbf{r}')\,\hat\Psi(\mathbf{r})}{|\mathbf{r}-\mathbf{r}'|},
\end{align}
where $\hat\Psi$ ($\hat\Psi^\dagger$) are the electronic annihilation (creation) operators. For the sake of compactness, we suppressed the electron spin index, whose effect will amount to an additional factor of 2 in electronic response functions. 
The first term represents the kinetic energy of free electrons with the (effective) mass~$m$. In the external potential $e\varphi_\mathrm{ext}(\mathbf{r})$ we explicitly separated the electron charge $e<0$.
The last term is the Coulomb interaction.
It is convenient to rewrite this Hamiltonan as
\begin{align}
\hat{H} = {}&{}\int d^3\mathbf{r}\left[\hat\Psi^\dagger(\mathbf{r})\left(-\frac{\hbar^2\nabla^2}{2m}\right)\hat\Psi(\mathbf{r}) + \varphi_\mathrm{ext}(\mathbf{r})\,\hat\rho(\mathbf{r})\right] + {}\nonumber\\
{}&{}+ \frac{1}{2}\int d^3\mathbf{r}\, d^3\mathbf{r}'\, 
\frac{:\hat\rho(\mathbf{r})\,\hat\rho(\mathbf{r}'):}{|\mathbf{r}-\mathbf{r}'|},
\label{eq:CoulombHamiltonian}
\end{align}
\end{subequations}
where $:\ldots:$ denotes the normal ordering of the electronic field operators,
\begin{equation}
:\hat\Psi^\dagger(\mathbf{r}_1)\hat\Psi(\mathbf{r}_2)\hat\Psi^\dagger(\mathbf{r}_3)\hat\Psi(\mathbf{r}_4):\;\equiv\hat\Psi^\dagger(\mathbf{r}_1)\hat\Psi^\dagger(\mathbf{r}_3)\hat\Psi(\mathbf{r}_4)\hat\Psi(\mathbf{r}_2),
\end{equation}
and we introduce the operators of the charge density and the electric current density,
\begin{subequations}\label{eq:rhorjr}
\begin{align}
&\hat\rho(\mathbf{r})=e\,\hat\Psi^\dagger(\mathbf{r})\,\hat\Psi(\mathbf{r}),\\
&\hat{\mathbf{j}}(\mathbf{r})\equiv\frac{ie\hbar}{2m}
\left[\left(\grad\hat\Psi^\dagger(\mathbf{r})\right)\hat\Psi(\mathbf{r})
-\hat\Psi^\dagger(\mathbf{r})\left(\grad\hat\Psi(\mathbf{r})\right)\right],
\end{align}\end{subequations}
with the summation over the spin index implied. We introduce two other operator fields,
\begin{subequations}\label{eq:Eb}
\begin{align}
&\hat{\mathbf{E}}(\mathbf{r})\equiv
-\grad\int\frac{\hat\rho(\mathbf{r}')}{|\mathbf{r}-\mathbf{r}'|}\,d^3\mathbf{r}',\\
&\hat{\mathbf{b}}(\mathbf{r})\equiv
\grad\times\int\frac{\hat{\mathbf{j}}(\mathbf{r}')}{|\mathbf{r}-\mathbf{r}'|}\,d^3\mathbf{r}',
\end{align}\end{subequations}
which satisfy the following equations:
\begin{subequations}\label{eq:Maxwell}
\begin{align}
&\grad\cdot\hat{\mathbf{E}}(\mathbf{r})=4\pi\hat\rho(\mathbf{r}),\\
&\grad\times\hat{\mathbf{E}}(\mathbf{r})=0,\label{eq:Faraday}\\
&\grad\cdot\hat{\mathbf{b}}(\mathbf{r})=0,\\
&\grad\times\hat{\mathbf{b}}(\mathbf{r})=4\pi\,\hat{\mathbf{j}}(\mathbf{r})+\frac{i}\hbar\left[\hat{H},\hat{\mathbf{E}}(\mathbf{r})\right].
\end{align}\end{subequations}
The first two equations are the Maxwell's equations for the electrostatic elecric field $\mathbf{E}$, while the last two equations are the Maxwell's equations for the magnetic field~$\mathbf{B}$, up to the factor $1/c$ on the right-hand side of the last equation (indeed, the last term with the commutator becomes $\partial\hat{\mathbf{E}}/\partial{t}$ in the Heisenberg representation). 

The appearance of the magnetic field may look somewhat surprising in our problem with purely electrostatic interactions. Strictly speaking, the Coulomb limit is obtained by sending the speed of light $c\to\infty$ in the Maxwell's equations, so the magnetic field $\mathbf{B}=O(1/c)$ vanishes, but our $\mathbf{b}=c\mathbf{B}$ stays finite. The only term in the Maxwell's equations that disappears at $c\to\infty$ is $-(1/c^2)\partial\hat{\mathbf{b}}/\partial{t}$, which would stand on the right-hand side of Eq.~(\ref{eq:Faraday}).

To define the energy current operator $\hat{\mathbf{Q}}(\mathbf{r})$, one must first define the energy density operator $\hat{\mathcal{H}}(\mathbf{r})$, such that the Hamiltonian $\hat{H}=\int{d}^3\mathbf{r}\,\hat{\mathcal{H}}(\mathbf{r})$, and then define an operator $\hat{\mathbf{Q}}(\mathbf{r})$ which satisfies the energy continuity equation,
\begin{equation}\label{eq:energy_continuity}
\frac{\partial\hat{\mathcal{H}}(\mathbf{r})}{\partial{t}}=\frac{i}\hbar\left[\hat{H},\hat{\mathcal{H}}(\mathbf{r})\right]=-\grad\cdot\hat{\mathbf{Q}}(\mathbf{r}).
\end{equation}
In principle, it is possible to find many operators $\hat{\mathcal{H}}(\mathbf{r})$ which give $\hat{H}=\int{d}^3\mathbf{r}\,\hat{\mathcal{H}}(\mathbf{r})$. The gauge-nvariant definition is~\cite{Catelani2005}
\begin{subequations}\begin{align}\label{eq:energy_density}
&\hat{\mathcal{H}}(\mathbf{r})=\hat{\mathcal{K}}(\mathbf{r})+\varphi_\mathrm{ext}(\mathbf{r})\,\hat\rho(\mathbf{r})
+\frac{:\hat{\mathbf{E}}(\mathbf{r})\cdot\hat{\mathbf{E}}(\mathbf{r}):}{8\pi},\\
&\hat{\mathcal{K}}(\mathbf{r})=\frac{\hbar^2}{2m}
\left(\grad\hat\Psi^\dagger(\mathbf{r})\right)\cdot
\left(\grad\hat\Psi(\mathbf{r})\right).
\end{align}\end{subequations}
Indeed, integrating this expression by parts and using $\nabla^2(1/r)=-4\pi\delta(\mathbf{r})$, we recover Hamiltonian~(\ref{eq:CoulombHamiltonian}).

Adopting definition (\ref{eq:energy_density}), after a somewhat tedious but straightforward evaluation of the commutator $[\hat{H},\hat{\mathcal{H}}(\mathbf{r})]$, we recover Eq.~(\ref{eq:energy_continuity}) with the energy current density defined as
\begin{align}
\hat{\mathbf{Q}}={}&{}\frac{i\hbar}{2m}\left[\left(\grad\hat\Psi^\dagger\right)
\left(-\frac{\hbar^2\nabla^2}{2m}\hat\Psi\right)
-\left(-\frac{\hbar^2\nabla^2}{2m}\hat\Psi^\dagger\right)
\left(\grad\hat\Psi\right)\right]\nonumber\\
{}&{}+\varphi_{\mathrm{ext}}\,\hat{\mathbf{j}}
+\frac{:\hat{\mathbf{E}}\times\hat{\mathbf{b}}:}{4\pi}.
\end{align}
Here the first line represents the kinetic energy current, the first term on the second line is the potential energy current, and the last term is the Poynting vector usually defined as $\mathbf{S}=c[\mathbf{E}\times\mathbf{B}]/(4\pi)$. Again, in our electrostatic problem the magnetic field $\mathbf{B}=O(1/c)\to0$, but the Poynting vector remains finite due to the extra factor of~$c$. Using the Maxwell's equations~(\ref{eq:Maxwell}), we obtain the standard relation
\begin{equation}\label{eq:divS}
\grad\cdot\,\hat{\mathbf{S}}(\mathbf{r})=-:\hat{\mathbf{j}}(\mathbf{r})\cdot\hat{\mathbf{E}}(\mathbf{r}):-\frac{i}\hbar
\left[\hat{H},\frac{:\hat{\mathbf{E}}(\mathbf{r})\cdot\hat{\mathbf{E}}(\mathbf{r}):}{8\pi}\right].
\end{equation}

Let us now consider an infinitely thin 2D layer, located at $z=0$. Then, the 3D current density (directed in the layer plane) $\hat{\mathbf{j}}(\mathbf{r})=\hat{\mathbf{j}}_{2D}(\mathbf{r}_\|)\,\delta(z)$, while $\hat{\mathbf{E}}(\mathbf{r}_\|,z=0)$ is finite (we denoted $\mathbf{r}_\|=(x,y)$ and introduced the 2D surface current density $\hat{\mathbf{j}}_{2D}$). Then, the normal component of the Poynting vector $\hat{S}_z$ has a discontinuity which is found by integrating the normal component of Eq.~(\ref{eq:divS}) over~$z$ between $z=0^-$ and $z=0^+$:
\begin{align}
\hat{J}(\mathbf{r}_\|){}&{}=\hat{S}_z(\mathbf{r}_\|,z=0^-)-\hat{S}_z(\mathbf{r}_\|,z=0^+)={}\nonumber\\
&{}=\,:\hat{\mathbf{j}}_{2D}(\mathbf{r}_\|)\cdot\hat{\mathbf{E}}(\mathbf{r}_\|,z=0):.
\end{align}
The left-hand side of this equation is the radiative power deposited at the point $\mathbf{r}_\|$ of the sample. The right-hand side is Eq.~(\ref{eqn:heatop}) of the main text.

We perform the last manipulation, replacing the normally ordered product by half of the anticommutatior,
\begin{align}
&:\hat{\mathbf{j}}_{2D}(\mathbf{r}_\|)\cdot\hat{\mathbf{E}}(\mathbf{r}_\|,0):\,
\to\nonumber\\
&{}\to\frac12\left\{\hat{\mathbf{j}}_{2D}(\mathbf{r}_\|)\cdot\hat{\mathbf{E}}(\mathbf{r}_\|,0)
+\hat{\mathbf{E}}(\mathbf{r}_\|,0)\cdot\hat{\mathbf{j}}_{2D}(\mathbf{r}_\|)\right\}.
\label{eq:normal_to_anticommutator}
\end{align}
The equivalence between the two expressions is obvious for the field produced by charges in the other layer, since the fermionic field operators trivially anticommute. For the charges in the same layer, let us write the surface current $\hat{\mathbf{j}}_{2D}(\mathbf{r}_\|)$ and density $\rho_{2D}(\mathbf{r}_\|)$ in the same form as Eqs.~(\ref{eq:rhorjr}), with the replacement $\mathbf{r}\to\mathbf{r}_\|$, $\grad\to\grad_\|$, and the electric field 
\begin{equation}
\hat{\mathbf{E}}(\mathbf{r}_\|,0)=-\grad_\|\int{v}(\mathbf{r}_\|-\mathbf{r}_\|')\,\hat\rho_{2D}(\mathbf{r}_\|')\,d^2\mathbf{r}_\|',
\end{equation}
where ${v}(\mathbf{r}_\|-\mathbf{r}_\|')$ is a regularized version of $1/|\mathbf{r}_\|-\mathbf{r}_\|'|$ due to integration of $1/|\mathbf{r}-\mathbf{r}'|$ over $z,z'$ within the layer. The difference between half the anticommutator and the normally ordered product is
\begin{align*}
&-\frac{ie^2\hbar}{4m}\int{d}^2\mathbf{r}_\|'\sum_k
\frac{\partial{v}(\mathbf{r}_\|-\mathbf{r}_\|')}{\partial{x}_k}\,
\delta(\mathbf{r}_\|-\mathbf{r}_\|')\times{}\\
&\qquad\times\left(\frac\partial{\partial{x}_k}-\frac\partial{\partial{x}_k'}\right)\left[
\hat\Psi^\dagger(\mathbf{r}_\|)\,\hat\Psi(\mathbf{r}_\|')
-\hat\Psi^\dagger(\mathbf{r}_\|')\,\hat\Psi(\mathbf{r}_\|)\right],
\end{align*}
where $k$ labels the two in-plane Cartesian components~$x,y$. In the integrand, the second line is even with respect to interchange $\mathbf{r}_\|\leftrightarrow\mathbf{r}_\|'$, while the first line is odd, due to the gradient of the even ${v}(\mathbf{r}_\|-\mathbf{r}_\|')$. As a result, the integral over $\mathbf{r}_\|'-\mathbf{r}_\|$ vanishes. Thus, in the following we will calculate the noise spectrum of the second line in Eq.~(\ref{eq:normal_to_anticommutator}). This quantity is easier to evaluate, since it amounts to a calculation of joint moments of currents and densities.

\subsection{Generating functional for density and current and random phase approximation}
\label{Ann:gen_func}

In our calculation, we handle the impurity scattering and the electron-electron interaction using the standard approximations. Namely, we work in the leading order in the weak disorder parameter $1/(p_F\ell)\ll1$, where $p_F$ and $\ell=v_F\tau$ are the electron Fermi momentum and the mean free path due to the impurity scattering. The electron-electron interaction is included within the random phase approximation (RPA), which is expected to be valid for high electron density systems such as the metals treated in this work; it is also equivalent to treating the radiative heat tranfer using linear electrodynamics and neglecting the nonlinear response of the material.

The operator average $\langle\{\hat{J}(\mathbf{r}_\|,t),\hat{J}(\mathbf{r}_\|',t')\}\rangle$ can be evaluated by summing up the RPA diagrammatic series in the operator formulation of the non-equilibrium (Keldysh) perturbation theory.
However, we found it technically more convenient to evaluate this average via the functional differentiation of the generation functional for currents and densities, obtained in the path integral approach for non-equilibrium systems~\cite{Kamenev2011}.
The generating functional, $\mathcal{Z}[\phi,\mathbf{a}]$, depends on two (scalar and vector) source fields, $\phi(\mathbf{r},t)$ and $\mathbf{a}(\mathbf{r},t)$, which are conjugate to the charge density and current. It is defined as the weighted functional field integral
\begin{equation}
\mathcal{Z}[\phi,\mathbf{a}] = \int \mathcal{D}[\bar{\psi},\psi] 
\exp\left[\frac{i}\hbar\int_\mathcal{C} dt\,(\mathcal{L}_0+\mathcal{L}_\mathrm{int}+\mathcal{L}_\mathrm{s})\right],
\end{equation}
where $\bar\psi(\mathbf{r},t)$ and $\psi(\mathbf{r},t)$ are independent Grassman variables necessary to describe a fermionic system. The integration in time is over the Keldysh contour, $\mathcal{C}$, which begins at $-\infty$, when the interactions are adiabatically switched on, runs forward to $+\infty$ and then runs backwards to $-\infty$, ensuring the property $\mathcal{Z}[0,0]=1$. The three contributions to the Lagrangian for interacting electrons, corresponding to the Hamiltonian \eqref{eq:CoulombHamiltonian} and including the source fields, are given explicitly by 
\begin{subequations}\label{eqn:App:action}\begin{align}
\mathcal{L}_0={}&{} \int_\mathbf{r} \, \bar{\psi}(\mathbf{r}, t)\left[ i\hbar\,\frac{\partial}{\partial{t}} + \frac{\hbar^2\nabla^2}{2m} - e\varphi_\mathrm{ext}(\mathbf{r})\right] \psi(\mathbf{r}, t), \\
\mathcal{L}_\mathrm{int}={}&{}-\frac{e^2}{2}\int_{\mathbf{r},\mathbf{r}'} \bar{\psi}(\mathbf{r},t)\,
\psi(\mathbf{r}, t)\, V_0(\mathbf{r},\mathbf{r}')\, \bar{\psi}(\mathbf{r}', t)\,\psi(\mathbf{r}', t) ,\\
\mathcal{L}_\mathrm{s}= {}&{}\frac{ie\hbar}{2m}\int_\mathbf{r}\,\bar{\psi}(\mathbf{r}, t)\left\{\mathbf{a}(\mathbf{r},t)\cdot\grad+\grad\cdot\mathbf{a}(\mathbf{r},t)\right\}\psi(\mathbf{r},t)-{}\nonumber\\
{}&{}-e\int_\mathbf{r}\bar{\psi}(\mathbf{r}, t)\,{\psi}(\mathbf{r}, t)\,\phi(\mathbf{r},t).
\end{align}\end{subequations}
where we introduced a short-hand notation for the spatial integral.

At this point, we do not have to specify the geometry of the system. Equations~(\ref{eqn:App:action}) can describe a three-dimensional system with $\mathbf{r}=(x,y,z)$, then $\int_\mathbf{r}\equiv\int{d}^3\mathbf{r}$, the Coulomb interaction potential in the second line is simply $V_0(\mathbf{r},\mathbf{r}')=1/|\mathbf{r}-\mathbf{r}'|$, and $\grad$ is the three-dimensional gradient operator, acting on all functions to the right of it. For two thin layers placed at $z=0$ and $z=d$, one can view $\mathbf{r}$ as a combination of the 2D position vector $\mathbf{r}_\|=(x,y)$ and the discrete layer index $\lambda=1,2$, so that $\int_\mathbf{r}=\int{d}^2\mathbf{r}_\|\sum_\lambda$. Then $\grad$ should be understood as the in-plane gradient $\grad_\|$, multiplied by the unit matrix in the layer space. The Coulomb interaction also becomes a matrix in the layer space:
\begin{equation}\label{eq:matrixCoulomb_r}
V_0(\mathbf{r},\mathbf{r}')=\begin{pmatrix}
1/|\mathbf{r}_\|-\mathbf{r}_\|'| & 1/\sqrt{|\mathbf{r}_\|-\mathbf{r}_\|'|^2+d^2} \\
1/\sqrt{|\mathbf{r}_\|-\mathbf{r}_\|'|^2+d^2} & 1/|\mathbf{r}_\|-\mathbf{r}_\|'|
\end{pmatrix}.
\end{equation}
The source fields $\phi(\mathbf{r},t)$ and $\mathbf{a}(\mathbf{r},t)$ must not be confused with the gauge scalar and vector potentials (in particular, there is no $\mathbf{a}^2$ in the Lagrangian). They serve only to extract the average values of the density and current as well as their higher moments, e.~g.
\begin{equation}
\langle\hat{\mathbf{j}}(\mathbf{r},t)\rangle=\pm{i}\hbar\left.\frac{\delta\mathcal{Z}}{\delta\mathbf{a}(\mathbf{r},t)}\right|_{\phi,\mathbf{a}=0},
\end{equation}
where the ``$+$'' or the ``$-$'' sign should be chosen for time located on the forward or the backward branch of the contour. The source field $\mathbf{a}(\mathbf{r}, t)$ may be a three-dimensional vector, or have only the two in-plane components for the case of two layers.

We follow the standard procedure described in Ref.~\cite{Kamenev2011} to perform the disorder averaging.  The terms which are quartic in fermionic fields, originating from the disorder averaging and Coulomb interaction are decoupled via the introduction of Hubbard-Stratonovich bosonic fields $Q$ and $\Phi$, respectively. The resulting integral over the fermionic fields becomes Gaussian and so may be performed exactly, resulting in an effective nonlinear bosonic theory. Restricting ourselves to the leading order in $1/(p_F\ell)$ and to the RPA, we set the $Q$~field to its saddle-point value and expand the action to second order in the $\Phi$~field and in the source fields, and perform the Gaussian integration over the $\Phi$ field. The resulting generating functional can be written as
\begin{equation}
\mathcal{Z}[\mathcal{A}]=\exp\left[-\frac{i}\hbar\int_{X,X'}
\mathcal{A}_k^\alpha(X)\,\tilde{\Pi}_{kk'}^{\alpha\beta}(X,X')\,\mathcal{A}^\beta_{k'}(X')\right],
\label{eqn:genfunc}
\end{equation}
where the 4-vectors $X=(\mathbf{r},t)$ and $\mathcal{A}=(\mathbf{a},\phi)$ are introduced for compactness, and their components are labeled by the subscripts~$k,k'$. The integration $\int_X=\int_\mathbf{r}\int_{-\infty}^\infty{d}t$, while the forward/backward structure of the time contour is taken care of by introduction of the classical/quantum components of the fields, $\mathcal{A}_k^\mathrm{cl}=(\mathcal{A}_k^\mathrm{f}+\mathcal{A}_k^\mathrm{b})/2$, $\mathcal{A}_k^\mathrm{q}=(\mathcal{A}_k^\mathrm{f}-\mathcal{A}_k^\mathrm{b})/2$, which are labelled by the Greek superscripts $\alpha,\beta=\mathrm{cl},\mathrm{q}$. Summation over all repeating indices is implied. Finally, $\tilde{\Pi}_{kk'}^{\alpha\beta}(X,X')$ is defined in the following section.

\subsection{Polarisation operator and screened interaction}

The main object appearing in Eq.~(\ref{eqn:genfunc}) is the RPA-dressed polarisation operator $\tilde{\Pi}_{kk'}^{\alpha\beta}(X,X')$, satisfying the integral equation that we write symbolically as
\begin{equation}\label{eq:Dyson}
\tilde\Pi=\Pi+\Pi{V}\tilde\Pi,
\end{equation}
where the products imply the convolution over space and time variables and the matrix product over all other indices.
The tildeless ${\Pi}_{kk'}^{\alpha\beta}(X,X')$ is the polarization operator of the non-interacting electronic system, which is nothing else but the disorder-averaged current and density pair correlation functions,
\begin{equation}
\Pi_{kk'}^{\alpha\beta} (X,X') = -\frac{2i}\hbar\, \sigma_1^{\alpha\alpha'}\langle\mathrm{T}_\mathcal{C} \hat{\mathcal{J}}_k^{\alpha'}(X)\,\hat{\mathcal{J}}_{k'}^{\beta'}(X')\rangle\sigma_1^{\beta'\beta}, \label{eqn:App:GFav}
\end{equation}
where $\hat{\mathcal{J}}=(\hat{\mathbf{j}},\hat\rho)$ is the four-vector corresponding to the observables, the current and the density, $\sigma_1$~is the first Pauli matrix in the $2\times2$ Keldysh space, the time dependence of the operators is determined by the non-interacting Hamiltonian, and $\mathrm{T}_\mathcal{C}$ denotes the time ordering along the Keldysh contour. 
The bare Coulomb interaction $V$ has only the density-density components,
\begin{equation}
V_{kk'}^{\alpha\beta}(X,X')=\delta_{k\rho}\delta_{k'\rho}\sigma_1^{\alpha\beta}\,
\delta(t-t')\,V_0(\mathbf{r},\mathbf{r}').
\end{equation}
Had we wanted to include the full retarded interaction via the electromagnetic field, $V$~would include the photonic propagator of the transverse vector potential in the Coulomb gauge.
Similarly to the RPA dressed polarisation operator, one can also introduce the RPA screened interaction which may itself be expressed via the iterative Dyson equation
\begin{equation}
\tilde{V}_{\rho\rho} = V_0\sigma_1 + V_0\sigma_1 \Pi_{\rho\rho}\tilde{V}_{\rho\rho},
\label{eqn:App:U_RPA_iter}
\end{equation}
so that one can write $\tilde\Pi=\Pi+\Pi\tilde{V}\Pi$.

The polarization operator has only three non-zero components, which can be explicitly written as
\begin{subequations}\begin{align}
&\begin{pmatrix}
\Pi_{kk'}^{\mathrm{cl,cl}} & \Pi_{kk'}^{\mathrm{cl,q}} \\ \Pi_{kk'}^{\mathrm{q,cl}} & \Pi_{kk'}^{\mathrm{q,q}}
\end{pmatrix}
=\begin{pmatrix}
0 & \Pi_{kk'}^A \\ \Pi_{kk'}^R & \Pi_{kk'}^K
\end{pmatrix},\\
&\Pi_{kk'}^R(X,X')=-\frac{i}\hbar\,\theta(t-t')\,\langle[\hat{\mathcal{J}}_k(X),\hat{\mathcal{J}}_{k'}(X')]\rangle,\label{eq:PiR=}\\
&\Pi_{kk'}^A(X,X')=\Pi_{k'k}^R(X',X),\label{eq:PiA=}\\
&\Pi_{kk'}^K(X,X')=-\frac{i}\hbar\,\langle\{\hat{\mathcal{J}}_k(X),\hat{\mathcal{J}}_{k'}(X')\}\rangle
=\Pi_{k'k}^K(X',X),
\end{align}\end{subequations}
where $[\ldots]$ and $\{\ldots\}$ denote the commutator and the anticommutator, respectively. The average is taken over the non-interacting non-equilibrium density matrix of the system where the two electronic subsystems under consideration (the two layers) are held at different temperatures $T_1,T_2$. In a steady state, the averages depend only on the time difference $t-t'$, so we can define the Fourier transform,
\begin{equation}
\Pi_{kk'}^{\alpha\beta}(X,X')=\int_{-\infty}^\infty\frac{d\omega}{2\pi}\,e^{-i\omega(t-t')}
\Pi_{kk'}^{\alpha\beta}(\mathbf{r},\mathbf{r}',\omega).
\end{equation}
In thermal equilibrium at temperature $T$ the three components of the polarization operator obey the relation~\cite{Kamenev2011}
\begin{equation}
\Pi_{kk'}^K(\mathbf{r},\mathbf{r}',\omega) = \left[ \Pi_{kk'}^R(\mathbf{r},\mathbf{r}',\omega)-\Pi_{kk'}^A(\mathbf{r},\mathbf{r}',\omega)\right]\coth\frac{\hbar\omega}{2T}, \label{eqn:AppGFK}
\end{equation}
valid both for the non-interacting and the RPA dressed polarization operator. In the problem of radiative heat transfer between two electronic subsystems, one assumes that electrons in each subsystem are in equilibrium among themselves, so Eq.~(\ref{eqn:AppGFK}) is assumed to be valid for the non-interacting polarization operators of each subsystem, each with its own temperature.

Different current and density components of the polarisation operator are related via the continuity equation,
\begin{equation}\label{eq:continuity}
\grad\cdot \hat{\mathbf{j}}(\mathbf{r}, t) + \frac{\partial \hat\rho(\mathbf{r}, t)}{\partial t} = 0,
\end{equation}
which leads to the relations:
\begin{subequations}\label{eqn:App:D_rho-j}\begin{align} 
\frac{\partial{\Pi}^{K}_{j_i j_k}(\mathbf{r}, \mathbf{r}', \omega)}{\partial{x}_k'}
&=-i\omega\,\Pi^K_{j_i\rho}(\mathbf{r}, \mathbf{r}', \omega),
\label{eqn:App:D_jrho-jj}\\
\frac{\partial{\Pi}^{K}_{j_i j_k}(\mathbf{r}, \mathbf{r}', \omega)}{\partial{x}_i}
&=i\omega\,\Pi^K_{\rho j_k}(\mathbf{r}, \mathbf{r}', \omega),
\label{eqn:App:D_rhoj-jj}\\
\frac{\partial^2\Pi^K_{j_i j_k}(\mathbf{r}, \mathbf{r}', \omega)}{\partial{x}_i\,\partial{x}_k'}
&=\omega^2\Pi^K_{\rho\rho}(\mathbf{r}, \mathbf{r}', \omega).
\label{eqn:App:D_rhorho-jj}
\end{align}\end{subequations}
The same relations hold for the difference $\Pi^R-\Pi^A$, but for $\Pi^R$ and $\Pi^A$ separately, the time derivative of the extra factor $\theta(t-t')$ produces an additional term:
\begin{subequations}\begin{align}
\frac{\partial\Pi^{R.A}_{j_i j_k}(\mathbf{r}, \mathbf{r}', \omega)}{\partial{x}_i}
={}&{} i\omega\Pi^{R,A}_{\rho{j}_k}(\mathbf{r}, \mathbf{r}', \omega)-
\frac{i}\hbar\langle[\hat\rho(\mathbf{r}),\hat{j}_k(\mathbf{r}')]\rangle,\\
\frac{\partial\Pi^{R.A}_{j_i j_k}(\mathbf{r}, \mathbf{r}', \omega)}{\partial{x}_k'}
={}&{} -i\omega\Pi^{R,A}_{{j}_i\rho}(\mathbf{r}, \mathbf{r}', \omega)+
\frac{i}\hbar\langle[\hat{j}_i(\mathbf{r}),\hat\rho(\mathbf{r}')]\rangle.
\end{align}\end{subequations}
Instead of these latter relations, we employ different ones which originate from the fact that expression~(\ref{eq:PiR=}) has the structure of the Kubo formula for the response of the charge density or current to an external scalar or vector potential. Then the continuity equation~(\ref{eq:continuity}) relates different components of $\Pi^R(\mathbf{r},\mathbf{r}',\omega)$ to the nonlocal conductivity tensor $\sigma_{ik}(\mathbf{r},\mathbf{r}',\omega)$, which gives the response of the current density to an external electric field (which can be introduced via scalar or vector potential, as $-\grad\varphi$ or $-(1/c)\partial\mathbf{A}/\partial{t}$):
\begin{subequations}\label{eq:PiR=sigma}\begin{align}
\Pi^{R}_{\rho{j}_k}(\mathbf{r}, \mathbf{r}', \omega)={}&{}
-\frac{\partial\sigma_{ik}(\mathbf{r}, \mathbf{r}', \omega)}{\partial{x}_i},\\
\Pi^{R}_{j_i\rho}(\mathbf{r}, \mathbf{r}', \omega)={}&{}
\frac{\partial\sigma_{ik}(\mathbf{r}, \mathbf{r}', \omega)}{\partial{x}_k'},
\label{eq:PiRirho=sigma}\\
i\omega\,\Pi^{R}_{\rho\rho}(\mathbf{r}, \mathbf{r}', \omega)={}&{}
\frac{\partial^2\sigma_{ik}(\mathbf{r}, \mathbf{r}', \omega)}{\partial{x}_i\,\partial{x}_k'}
,\label{eq:PiRrhorho=sigma}\\
\Pi^{R}_{j_ij_k}(\mathbf{r}, \mathbf{r}', \omega)={}&{}
-i\omega\,\sigma_{ik}(\mathbf{r}, \mathbf{r}', \omega)-{}\nonumber\\
&{}-\delta_{ik}\delta(\mathbf{r}-\mathbf{r}')\langle\hat\rho(\mathbf{r})\rangle\,\frac{e}m,\label{eq:PiRik=sigma}
\end{align}\end{subequations}
where the last term comes the diamagnetic contribution to the electric current, proportional to the vector potential.
By virtue or Eq.~(\ref{eq:PiA=}) and of the fact that a physical response function in the time representation must be real, the analogous relations for $\Pi^A$ read
\begin{subequations}\label{eq:PiA=sigma}\begin{align}
\Pi^{A}_{\rho{j}_k}(\mathbf{r}, \mathbf{r}', \omega)={}&{}
\frac{\partial\sigma_{ki}^*(\mathbf{r}', \mathbf{r}, \omega)}{\partial{x}_i},
\label{eq:PiArhok=sigma}\\
\Pi^{A}_{j_i\rho}(\mathbf{r}, \mathbf{r}', \omega)={}&{}
-\frac{\partial\sigma_{ki}^*(\mathbf{r}', \mathbf{r}, \omega)}{\partial{x}_k'},\\
-i\omega\,\Pi^{A}_{\rho\rho}(\mathbf{r}, \mathbf{r}', \omega)={}&{}
\frac{\partial^2\sigma_{ki}^*(\mathbf{r}', \mathbf{r}, \omega)}{\partial{x}_i\,\partial{x}_k'},\\
\Pi^{A}_{j_ij_k}(\mathbf{r}, \mathbf{r}', \omega)={}&{}
i\omega\,\sigma_{ki}^*(\mathbf{r}', \mathbf{r}, \omega)-{}\nonumber\\
&{}-\delta_{ik}\delta(\mathbf{r}-\mathbf{r}')\langle\hat\rho(\mathbf{r})\rangle\,\frac{e}m.
\label{eq:PiAik=sigma}
\end{align}\end{subequations}
%

\begin{widetext}
Resolving explicitly the $2\times2$ Keldysh space matrix structire of Eq.~(\ref{eq:Dyson}), we represent the dressed polarisation operator entirely in terms of the bare one and bare interactions:
\begin{align}
&\begin{pmatrix}
0 & \tilde{\Pi}^A \\ \tilde{\Pi}^R & \tilde{\Pi}^K 
\end{pmatrix}
=  \begin{pmatrix}
0 & \Pi^A(\openone-V_0\Pi^A)^{-1} \\ 
\Pi^R(\openone-V_0\Pi^R)^{-1} & (\openone-\Pi^RV_0)^{-1}\Pi^K(\openone-V_0\Pi^A)^{-1} 
\end{pmatrix}, \label{eqn:App:tildeD_rhorho}
\end{align}
where intermediate variables are integrated over and $\openone = \delta(\mathbf{r}-\mathbf{r}')$ is the Dirac delta function. 
Similarly, the solution of Eq.~(\ref{eqn:App:U_RPA_iter}) can be written as
\begin{align}
\begin{pmatrix}
 \tilde{V}_{\rho\rho}^K & \tilde{V}_{\rho\rho}^R \\  \tilde{V}_{\rho\rho}^A & 0 
\end{pmatrix}
= \begin{pmatrix}
(\openone-V_0\Pi_{\rho\rho}^R)^{-1}V_0\Pi_{\rho\rho}^K(\openone-V_0\Pi_{\rho\rho}^A)^{-1}V_0 & (\openone-V_0\Pi_{\rho\rho}^R)^{-1}V_0 \\ 
(\openone-V_0\Pi_{\rho\rho}^A)^{-1}V_0 & 0 
\end{pmatrix}. \label{eqn:App:U_RPA_matrix}
\end{align}
\end{widetext}

\subsection{Fluctuations of the heat current}
\label{sec:Appfluc}

Due to the Gaussian form of the generating functional~(\ref{eqn:genfunc}), evaluation of contour-ordered moments of currents and densities by functional differentiation is equivalent to applying Wick theorem on the currents' and densities' operators as if they were linear combinations of bosonic creation and annihilation operators for a collection of harmonic oscillators with pair averages
\begin{subequations}\label{eq:Wickdensity}\begin{align}
&\langle\{\hat{\mathcal{J}}_k(X),\hat{\mathcal{J}}_{k'}(X')\}\rangle=
i\hbar\tilde\Pi_{kk'}^K(X,X'),\\
&\langle[\hat{\mathcal{J}}_k(X),\hat{\mathcal{J}}_{k'}(X')]\rangle=
i\hbar\left[\tilde\Pi_{kk'}^R(X,X')-\tilde\Pi_{kk'}^A(X,X')\right].
\end{align}\end{subequations}
The deep reason for this analogy is the harmonic nature of the RPA approximation or of the linear electrodynamics, which treat the interacting electron as a harmonic polarisable medium. From now on, the calculation is quite analogous to that for the harmonic quantum circuit, presented in Appendix~\ref{app:circuit}.

Writing the heat current correlator as
\begin{align}
\frac{\langle \{\hat{J}(\mathbf{r}, t), \hat{J}(\mathbf{r}', t')\}\rangle}2
=\frac18\int_{\mathbf{r}_1,\mathbf{r}_2}
\frac{\partial{V}_0(\mathbf{r},\mathbf{r}_1)}{\partial{x}_i}\,
\frac{\partial{V}_0(\mathbf{r}',\mathbf{r}_2)}{\partial{x}_k'}\times{}\nonumber\\
\times\left\langle\left\{\{\hat\rho(\mathbf{r}_1,t),\hat{j}_i(\mathbf{r},t)\},
\{\hat\rho(\mathbf{r}_2,t'),\hat{j}_k(\mathbf{r}',t')\}\right\}\right\rangle,
\end{align}
applying the Wick theorem with the pair averages~(\ref{eq:Wickdensity}), and dropping the product of the average heat currents, $\langle\hat{J}\rangle^2$, we obtain the following expression for the heat current noise correlator
\begin{align}
\mathcal{S}(X,X')={}&-\frac{\hbar^2}4\int_{\mathbf{r}_1,\mathbf{r}_2}
\frac{\partial{V}_0(\mathbf{r},\mathbf{r}_1)}{\partial{x}_i}\,
\frac{\partial{V}_0(\mathbf{r}',\mathbf{r}_2)}{\partial{x}_k'}\times{}\nonumber\\
{}&{}\times
\left[\tilde\Pi^K_{\rho\rho}(\mathbf{r}_1,t,\mathbf{r}_2,t')\,\tilde\Pi^K_{j_ij_k}(\mathbf{r},t,\mathbf{r}',t')\right.+{}\nonumber\\
&\quad{}+\left.\tilde\Pi^K_{\rho{j}_k}(\mathbf{r}_1,t,\mathbf{r}',t')\,\tilde\Pi^K_{j_i\rho}(\mathbf{r},t,\mathbf{r}_2,t')\right.+{}\nonumber\\
&\quad{}+\left.\left(\tilde\Pi^K\to\tilde\Pi^R-\tilde\Pi^A\right)\right].
\label{eqn:App:S_Dtilde}
\end{align}

Our goal now is to transform Eq.~(\ref{eqn:App:S_Dtilde}), expressing the dressed polarisation operators in terms of the bare ones and the dressed interaction using Eqs.~(\ref{eqn:App:tildeD_rhorho}) and~(\ref{eqn:App:U_RPA_matrix}). Instead of $\tilde{V}_{\rho\rho}$, it turns out to be more convenient to pass to the RPA-dressed dipole-dipole interaction,
\begin{equation}
\tilde{D}^{R,A}_{ik}(\mathbf{r},\mathbf{r}',\omega)\equiv
-\frac{\partial^2\tilde{V}_{\rho\rho}^{R.A}(\mathbf{r}, \mathbf{r}', \omega)}{\partial{x}_i\,\partial{x}_k'}.
\end{equation}
Then, taking the first term in the square brackets in Eq.~(\ref{eqn:App:S_Dtilde}), we transform its first factor using Eq.~(\ref{eqn:App:D_rhorho-jj}):
\begin{align}
&\int_{\mathbf{r}_1,\mathbf{r}_2}
\frac{\partial{V}_0(\mathbf{r},\mathbf{r}_1)}{\partial{x}_i}\,
\frac{\partial{V}_0(\mathbf{r}',\mathbf{r}_2)}{\partial{x}_k'}\,
\tilde\Pi^K_{\rho\rho}(\mathbf{r}_1,\mathbf{r}_2,\omega)={}\nonumber\\
&{}=\frac{1}{\omega^2}\int_{\mathbf{r}_1,\mathbf{r}_2}\tilde{D}^R_{il}(\mathbf{r},\mathbf{r}_1,\omega)
\,\Pi^K_{j_lj_m}(\mathbf{r}_1,\mathbf{r}_2,\omega)\,\tilde{D}^A_{mk}(\mathbf{r}_2,\mathbf{r}',\omega).
\label{eq:VPiV_to_DPiD}
\end{align}
In the second factor we use $\tilde\Pi_{j_ij_k}=\Pi_{j_ij_k}+\Pi_{j_i\rho}\tilde{V}_{\rho\rho}\Pi_{\rho{j}_k}$, $\tilde{V}^K_{\rho\rho}=\tilde{V}^R_{\rho\rho}\Pi^K_{\rho\rho}\tilde{V}^A_{\rho\rho}$, and Eqs.~(\ref{eqn:App:D_rho-j}), (\ref{eq:PiRirho=sigma}), (\ref{eq:PiArhok=sigma}) to obtain
\begin{align}
\tilde\Pi^K_{j_ij_k}(\mathbf{r},\mathbf{r}',\omega)={}&{}
\left(\openone\delta_{il} - \frac{\sigma_{ip}\tilde{D}^R_{pl}}{i\omega}\right)\Pi^K_{j_lj_m} 
\times{}\nonumber\\ &{}\times
\left(\openone\delta_{mk}+\frac{\tilde{D}^A_{mq}(\sigma^\dagger)_{qk}}{i\omega}\right),
\label{eq:Pi_to_PiDPiDPi}
\end{align}
where all intermediate spatial variables are integrated over, and summation over repeating Cartesian indices is implied, similarly to the right-hand side of Eq.~(\ref{eq:VPiV_to_DPiD}). In addition to the conductivity $\sigma_{ik}(\mathbf{r},\mathbf{r}',\omega)$, we introducd a short-hand notation for its Hermitian conjugate, $(\sigma^\dagger)_{ik}(\mathbf{r},\mathbf{r}',\omega)=\sigma_{ki}^*(\mathbf{r}',\mathbf{r},\omega)$.  It is easy to check that the same relations~(\ref{eq:VPiV_to_DPiD}) and~(\ref{eq:Pi_to_PiDPiDPi}) hold if one replaces $\tilde\Pi^K_{\rho\rho}\to\tilde\Pi^R_{\rho\rho}-\tilde\Pi^R_{\rho\rho}$ on the left-hand side and $\Pi^K_{j_lj_m}\to\Pi^R_{j_lj_m}-\Pi^A_{j_lj_m}$ on the right-hand side.

The second term in the square brackets in Eq.~\eqref{eqn:App:S_Dtilde} is approached in a completely analogous way, eventually leading to the following expression for the noise correlator whose structure is very similar to that of Eq.~(\ref{eq:XiPiXi}):
\begin{subequations}\label{eqn:App:Stotal}\begin{align}
\mathcal{S}(X,X')={}&-\frac{\hbar^2}4 \int_{-\infty}^\infty \frac{d\omega_1\,d\omega_2}{(2\pi)^2}\,e^{i(\omega_1+\omega_2)(t-t')} \times{}\nonumber\\
&\times\Big[\Lambda^{K\rho\rho}_{ik}(\mathbf{r},\mathbf{r}',\omega_1)\,\Lambda^{Kjj}_{ik}(\mathbf{r},\mathbf{r}',\omega_2)+{}\nonumber\\
&\quad{}+\Lambda^{K\rho{j}}_{ik}(\mathbf{r},\mathbf{r}',\omega_1)\,\Lambda^{Kj\rho}_{ik}(\mathbf{r},\mathbf{r}',\omega_2)+{}\nonumber\\
&\quad{}+\Lambda^{RA\rho\rho}_{ik}(\mathbf{r},\mathbf{r}',\omega_1)\,\Lambda^{RAjj}_{ik}(\mathbf{r},\mathbf{r}',\omega_2)+{}\nonumber\\
&\quad{}+\Lambda^{RA\rho{j}}_{ik}(\mathbf{r},\mathbf{r}',\omega_1)\,\Lambda^{RAj\rho}_{ik}(\mathbf{r},\mathbf{r}',\omega_2)\Big],
\label{eq:S=LambdaLambda}
\end{align}
where we denoted (implying convolution over all intermediate spatial variables and summation over repeating Cartesian indices)
\begin{align}
&\Lambda^{K\rho\rho}=-\frac{\tilde{D}^R}{i\omega_1}\,\Pi^K\frac{\tilde{D}^A}{i\omega_1},
\label{eq:Lambdarhorho}\\
&\Lambda^{Kjj}=\left(\openone - \frac{\sigma\tilde{D}^R}{i\omega_2}\right)\Pi^K 
\left(\openone+\frac{\tilde{D}^A\sigma^\dagger}{i\omega_2}\right),\\
&\Lambda^{K\rho{j}}=-\frac{\tilde{D}^R}{i\omega_1}\,\Pi^K\left(\openone+\frac{\tilde{D}^A\sigma^\dagger}{i\omega_1}\right),\\
&\Lambda^{Kj\rho}=\left(\openone - \frac{\sigma\tilde{D}^R}{i\omega_2}\right)\Pi^K\,\frac{\tilde{D}^A}{i\omega_2},
\label{eq:Lambdajj}
\end{align}\end{subequations}
and $\Lambda^{RA}$ are defined in the same way with the replacement $\Pi^K\to\Pi^R-\Pi^A$.
All polarisation operators appearing here are current-current ones, so they carry Cartesian indices. The difference $\Pi^R-\Pi^A$ can be expressed via the conductivity using Eqs.~(\ref{eq:PiRik=sigma}) and (\ref{eq:PiAik=sigma}). The $\Pi^K$ component is also expressed in terms of the conductivity using Eq.~(\ref{eqn:AppGFK}) with different temperatures in each body. 

Even though our derivation was based on the model where only the Coulomb interaction between the electrons was included, Eqs.~(\ref{eqn:App:Stotal}) remain valid also for the full retarded interaction via the electromagnetic field, if one understands $\tilde{D}^{R}_{ik}(\mathbf{r},\mathbf{r}',\omega)$ as the dressed propagator of the electric field, which determines the response of the electric field $\mathbf{E}$ to an external polarisation $\mathbf{P}^\mathrm{ext}$,
\begin{equation}
E_i(\mathbf{r},\omega)=\int_{\mathbf{r}'}\tilde{D}^R_{ik}(\mathbf{r},\mathbf{r}',\omega)\,P_k^\mathrm{ext}(\mathbf{r}',\omega),
\end{equation}
found from the full set of Maxwell's equations in the presence of the medium whose response is characterised by the conductivity $\sigma_{ik}(\mathbf{r},\mathbf{r}',\omega)$:
\begin{subequations}\label{eq:MaxwellE}\begin{align}
&\frac{\omega^2}{c^2}\,\mathbf{E}-\grad\times\grad\times\mathbf{E}
+\frac{4\pi{i}\omega}{c^2}\,\mathbf{j}=-\frac{4\pi\omega^2}{c^2}\,\mathbf{P}^\mathrm{ext},\\
&j_i(\mathbf{r},\omega)=\int_{\mathbf{r}'}\sigma_{ik}(\mathbf{r},\mathbf{r}',\omega)\,E_k(\mathbf{r}',\omega).
\end{align}\end{subequations}


\subsection{Two-dimensional metallic layers}

Let us now focus on the specific geometry of two parallel identical two-dimensional metallic layers, separated by a distance~$d$ and characterised by the local Drude conductivity in the plane, $\sigma(\omega)=\sigma_\mathrm{dc}/(1-i\omega\tau)$, with a temperature-independent relaxation time~$\tau$. Since the system is translationally invariant in the $(x,y)$ plane, it is convenient to do the Fourier transform with respect to the in-plane coordinate difference $\mathbf{r}_\|-\mathbf{r}_\|'$, so that Eq.~(\ref{eq:PiRrhorho=sigma}) gives
\begin{equation}
\Pi^R_{\rho\rho}(\mathbf{r}_\|,\mathbf{r}_\|',\omega)=
\int\frac{d^2\mathbf{k}}{(2\pi)^2}\,e^{i\mathbf{k}(\mathbf{r}_\|-\mathbf{r}_\|')}\,\frac{k^2}{i\omega}\,\sigma(\omega)
\end{equation}
for each of the two layers. The Fourier transform of Eq.~(\ref{eq:matrixCoulomb_r}) is
\begin{equation}
V_0(\mathbf{k})=\frac{2\pi}k\begin{pmatrix} 1 & e^{-kd} \\ e^{-kd} & 1\end{pmatrix},
\end{equation}
so the screened dipole-dipole interaction is
\begin{align}
\tilde{D}_{ij}^R(\mathbf{k})={}&-\frac{2\pi{k}_ik_j/k}{(1-\zeta)^2-\zeta^2e^{-2kd}}\times{}\nonumber\\
&\times\begin{pmatrix} 1-\zeta(1-e^{-2kd}) & e^{-kd} \\
e^{-kd} & 1-\zeta(1-e^{-2kd})\end{pmatrix},
\end{align}
where we denoted $\zeta\equiv-2\pi{i}k\sigma(\omega)/\omega$ for compactness. Note that this expression coincides with the $c\to\infty$ limit of the electric field propagator obtained by solving the full set of Maxwell's equations or, equivalently, Eqs.~(\ref{eq:MaxwellE}), with the external in-plane polarisation $\mathbf{P}^\mathrm{ext}(\mathbf{r})=[\mathbf{P}_1\delta(z)+\mathbf{P}_2\delta(z-d)]e^{i\mathbf{k}\mathbf{r}_\|}$ and the 3D conductivity
\begin{align}
\sigma_{ij}(\mathbf{r},\mathbf{r}',\omega)={}&\sigma(\omega)\,\delta_{ij}\,\delta(\mathbf{r}_\|-\mathbf{r}_\|')\times{}\nonumber\\
&{}\times[\delta(z)\delta(z')+\delta(z-d)\,\delta(z'-d)].
\end{align}
Namely, the solution for the electric field propagator is
\begin{widetext}
\begin{align}
\tilde{D}^R_{ij}(\mathbf{k},\omega)={}&{}\frac{2\pi{i}(\omega/c)^2(\delta_{ij}-k_ik_j/k^2)}{[q+2\pi\omega\sigma(\omega)/c^2]^2-[2\pi\omega\sigma(\omega)/c^2]^2e^{2iqd}}
\begin{pmatrix}
q+[2\pi\omega(\omega)\sigma/c^2](1-e^{2iqd}) & qe^{iqd} \\ qe^{iqd} & q+[2\pi\omega\sigma(\omega)/c^2](1-e^{2iqd})
\end{pmatrix}
+{}\nonumber\\
{}&{}+\frac{2\pi{i}q(k_ik_j/k^2)}{[1+2\pi{q}\sigma(\omega)/\omega]^2-[2\pi{q}\sigma(\omega)/\omega]^2e^{2iqd}}
\begin{pmatrix}
1+[2\pi{q}\sigma(\omega)/\omega](1-e^{2iqd}) & e^{iqd} \\ e^{iqd} & 1+[2\pi{q}\sigma(\omega)/\omega](1-e^{2iqd})
\end{pmatrix},
\end{align}
\end{widetext}
where we denoted $q=\sqrt{\omega^2/c^2-k^2}\,\sign\omega$ if $|\omega|>ck$ and $q=i\sqrt{k^2-\omega^2/c^2}$ if $|\omega|<ck$.

In Eqs.~(\ref{eqn:App:Stotal}), $\mathbf{r}$ is now understood as the combination of the in-plane $\mathbf{r}_\|$ and the layer index, so all convolutions over $\mathbf{r}_\|$ become simple products of $2\times2$ matrices in the Fourier space. In particular,
\begin{align}
\Pi^K_{j_lj_m}(\mathbf{k},\omega)={}&-2i\omega\delta_{lm}\Re\sigma(\omega)\times{}\nonumber\\
&\times\begin{pmatrix}
\coth[\hbar\omega/(2T_1)] & 0 \\ 0 & \coth[\hbar\omega/(2T_2)]
\end{pmatrix}.\label{eq:PiK=sigma}
\end{align}
Representing the noise correlator in terms of its spectrum,
\begin{equation}
\mathcal{S}(X,X')=\int\frac{d^2\mathbf{K}}{(2\pi)^2}\,\frac{d\Omega}{2\pi}\,
e^{i\mathbf{K}(\mathbf{r}_\|-\mathbf{r}_\|')-i\Omega(t-t')}
\mathcal{S}(\mathbf{K},\Omega),
\end{equation}
the latter has the following structure:
\begin{align}
\mathcal{S}(\mathbf{K},\Omega)={}&-\frac{\hbar^2}4\int\frac{d^2\mathbf{k}}{(2\pi)^2}\,\frac{d\omega}{2\pi}\times{}\nonumber\\
&\times\left[\Lambda^{K\rho\rho}_{lm}(\mathbf{k},\omega)\,\Lambda^{Kjj}_{lm}(\mathbf{K}-\mathbf{k},\Omega-\omega)+\ldots\right],
\end{align}
where ``$\ldots$'' denotes the last three lines of Eq.~(\ref{eq:S=LambdaLambda}) with the same wave vector and frequency arguments. Each $\Lambda$ factor remains a $2\times2$ matrix in the in-plane Cartesian components $l,m=x,y$, which are summed over, and a $2\times2$ matrix in the layer index. Since we are looking at the noise of the power dissipated in layer~1, the layer index of both $\mathbf{r}$ and $\mathbf{r}'$ corresponds to layer~1. Furthermore, each $\Lambda$~factor contains either one diagonal matrix $\Pi^K$, given by Eq.~(\ref{eq:PiAik=sigma}), or $\Pi^R-\Pi^A$, given by the same equation but with the replacement $\coth\to1$. As a result, the noise spectrum will contain four terms, proportional to
\begin{align*}
&\coth\frac{\hbar\omega}{2T_1}\coth\frac{\hbar(\Omega-\omega)}{2T_1}+1,\\
&\coth\frac{\hbar\omega}{2T_1}\coth\frac{\hbar(\Omega-\omega)}{2T_2}+1,\\
&\coth\frac{\hbar\omega}{2T_2}\coth\frac{\hbar(\Omega-\omega)}{2T_1}+1,\\
&\coth\frac{\hbar\omega}{2T_2}\coth\frac{\hbar(\Omega-\omega)}{2T_2}+1.
\end{align*}
We are interested in the case $T_2\gg{T}_1$, so the last line gives the dominant contribution. This contribution contains the off-diagonal (in the layer index) matrix elements of the factors surrounding $\Pi^K$ in Eqs.~(\ref{eq:Lambdarhorho})--(\ref{eq:Lambdajj}), so $\openone$ does not contribute. Collecting all factors, we arrive at Eq.~(\ref{eq:SKOmega=}) of the main text.


\subsection{Plasmon resonance feature in the noise spectrum}
\label{ssec:S_plasmon}

Let us write $u_{12}(\mathbf{k},\omega)$ from Eq.~(\ref{eq:u12=}) of the main text as
\begin{align}
&u_{12}(\mathbf{k},\omega)=-\frac{2\pi{i}\omega{k}e^{-kd}(1-i\omega\tau)^2/\tau^2}%
{[\omega^2-\omega_-^2(k)+i\omega/\tau][\omega^2-\omega_+^2(k)+i\omega/\tau]},
\end{align}
where the symmetric and the antisymmetric plasmon frequencies are given by $\omega_\pm^2(k)=2\pi{k}(1\pm{e}^{-kd})(\sigma_\mathrm{dc}/\tau)$. Then Eq.~(\ref{eq:SKOmega=}) can be written as
\begin{align}
\mathcal{S}(\mathbf{K},\Omega)={}&\frac{4+\Omega^2\tau^2}2\int
\frac{d^2\mathbf{k}_1\,d^2\mathbf{k}_2\,d\omega_1\,d\omega_2}{(2\pi)^3}
\times{}\nonumber\\ &{}\times
\delta(\mathbf{k}_1+\mathbf{k}_2-\mathbf{K})\,\delta(\omega_1+\omega_2-\Omega)
\times{}\nonumber\\ &{}\times
(2\pi\sigma_\mathrm{dc})^4(\mathbf{k}_1\cdot\mathbf{k}_2)^2e^{-2k_1d-2k_2d}
\times{}\nonumber\\ &{}\times
\hbar^2\omega_1^3\omega_2^3\left(1+\coth\frac{\hbar\omega_1}{T_2}\coth\frac{\hbar\omega_2}{T_2}\right)
\times{}\nonumber\\ &{}\times
\prod_{j=1,2}\prod_{s=\pm}\frac1{[\omega_j^2-\omega_s^2(k_j)]^2\tau^2+\omega_j^2}.
\end{align}
Here we noted that the since the integration variables $\omega_1,\omega_2$ can be interchanged, one can symmetrize
\begin{align}
&|\sigma(\omega_1)|^2+\sigma(\omega_1)\,\sigma^*(\omega_2)\to\nonumber\\
&\to\frac{|\sigma(\omega_1)|^2+|\sigma(\omega_2)|^2
+\sigma(\omega_1)\,\sigma^*(\omega_2)+\sigma^*(\omega_2)\,\sigma(\omega_2)}2\nonumber\\
&{}=\frac{\sigma_\mathrm{dc}^2}2\,\frac{4+(\omega_1+\omega_2)^2\tau^2}{(1+\omega_1^2\tau^2)(1+\omega_2^2\tau^2)}.
\end{align}
We focus on the regime when the heat transfer is dominated by the antisymmetric plasmon resonance, that is $\hbar/\tau\ll{T}_2\ll{v}_-/d$, where $v_-=\sqrt{2\pi\sigma_\mathrm{dc}d/\tau}$ is the antisymmetric plasmon velocity~\cite{Wise2020}. The typical plasmon frequencies and wave vectors that contribute to $\langle{J}\rangle$ are $\omega\sim{T}_2/\hbar\gg1/\tau$, $k\sim{T}_2/(\hbar{v}_-)\ll1/d$. For the noise spectrum, we are interested in $\mathbf{K},\Omega$ such that  $|\Omega|\sim1/\tau\ll|\omega_{1,2}|\sim{T}_2/\hbar$ and $K\sim1/(v_-\tau)\ll{k}_{1,2}$. Then one can set $\omega_1=-\omega_2$ and $\mathbf{k}_1=-\mathbf{k}_2$ everywhere except the resonant denominators, which gives
\begin{align}
\mathcal{S}(\mathbf{K},\Omega)={}&\frac{4+\Omega^2\tau^2}{16}\int_0^\infty\frac{d\omega}{2\pi}\,
\frac{\hbar^2\omega^2}{\sinh^2[\hbar\omega/(2T_2)]}
\times{}\nonumber\\ &{}\times
\int\frac{d^2\mathbf{k}}{(2\pi)^2}\,
\frac{1}{4\tau^2(\omega-v_-k)^2+1}
\times{}\nonumber\\ &{}\times
\frac{1}{4\tau^2(\omega-\Omega-v_-|\mathbf{k}-\mathbf{K}|)^2+1}.
\end{align}
We represent the $\mathbf{k}$ integral in the polar coordinates $(k,\phi)$, approximating $|\mathbf{k}-\mathbf{K}| \approx k-K\cos\phi$. We first integrate over~$k$, replacing $k=\omega/v_-$ in the numerator and extending the integration limits to $-\infty<k<\infty$. The remaining $\omega$ and $\phi$ integrals separate and are calculated exactly, yielding Eq.~(\ref{eq:2d_res}) of the main text.

\bibliography{noise_paper}

\end{document}